\documentclass[aps,amsfonts,draft]{revtex4}

\usepackage{bm}

\tighten
\begin{document}

\preprint{APS/123-QED}

\title{Fine-structure constant variability, equivalence principle and
cosmology}

\author{Jacob D. Bekenstein}\email{bekenste@vms.huji.ac.il}
\homepage{http://www.phys.huji.ac.il/~bekenste/}
\affiliation{Racah Institute of Physics, Hebrew University of
Jerusalem\\ Givat Ram, Jerusalem 91904 ISRAEL}

\date{\today}

\begin{abstract}  It has been widely believed that variability of the
fine-structure constant $\alpha$ would imply detectable violations of
the weak equivalence principle.  This belief is not justified in
general.  It is put to rest here in the context of the general
framework for $\alpha$ variability [J. D. Bekenstein, Phys. Rev. D {\bf
25}, 1527 (1982)] in which the exponent of a scalar field plays the
role of the permittivity and inverse permeability of the vacuum.  The
coupling of particles to the scalar field is necessarily such that the
anomalous force acting on a charged particle by virtue of its mass's
dependence on the scalar field is cancelled by terms modifying the
usual Coulomb force.  As a consequence a particle's acceleration in
external fields depends only on its charge to mass ratio, in accordance
with the principle.  And the center of mass acceleration of a composite
object can be proved to be independent of the object's internal
constitution, as the weak equivalence principle requires.  Likewise the
widely  employed assumption that the Coulomb energy of matter is the
principal source of the scalar field proves wrong; Coulomb energy
effectively cancels out in the continuum description of the scalar
field's dynamics.  This cancellation resolves a cosmological conundrum:
with Coulomb energy as the source of the scalar field, the framework would
predict a decrease of
$\alpha$ with cosmological expansion, whereas an increase is claimed to
be observed.   Because of the said cancellation, magnetic energy of
cosmological baryonic matter is the main source of the scalar field.
Consequently the expansion is accompanied by an increase in $\alpha$;
for reasonable values of the framework's sole parameter, this occurs at
a rate consistent with the observers' claims.

\end{abstract}


\maketitle

\section{\label{sec:intro}Introduction}

Observations of fine structure multiplet splittings in the absorption
line systems of distant quasars have lately suggested~\cite{webb} that
the fine-structure constant $\alpha$ at cosmological epochs with redshift
1--3.5  was lower than it is today.  This evidently revolutionary claim
begs for a theoretical framework to enable judgment as to whether such a
variation can be consistent with accepted physical principles.  Variation
of
$\alpha$ was first considered theoretically by Jordan~\cite{jordan},
Teller~\cite{teller} and Stanyukovich~\cite{stan}.

Already before Gamow's influential speculation that
$\alpha$ varies linearly with cosmological time~\cite{gamow}, Dicke had
made the point that any variation of $\alpha$ can be regarded equally
well as due to variation of particle charge, or alternatively of the
speed of light $c$ or of
$\hbar$, with the choice being a matter of convenience, not
physics~\cite{dicke}.  He also exhibited a theory in which the Maxwell
invariant couples linearly to a standard massless scalar field as an
example of a variable
$\alpha$ theory.  In this theory the permittivity and reciprocal
permeability of the vacuum vary in consonance with the scalar field; it
can also be regarded as a variable charge theory.  Independently of the
specific theory of $\alpha$ variation, Dicke provided an
argument~\cite{dicke,dicke2} that spatial variation of $\alpha$, which
would be expected to accompany cosmological temporal variation if the
underlying theory is covariant, contradicts the weak equivalence
principle (WEP).  The essence of the argument is that a non-negligible
fraction $\zeta$ of the mass $M$ of any chunk of ordinary neutral
matter is Coulombic in origin.  The Coulomb energy $E_C=\zeta Mc^2$
should scale with the square of the constituent charges, meaning it
should be proportional to $\alpha$, and should thus depend on position
if $\alpha$ does.  The force on mass $M$ should thus contain in addition
to the Newtonian part $-M\bm{\nabla}\phi_N$ an anomalous portion
$-\bm{\nabla}E_C=-(\partial E_C/\partial
\alpha)\bm{\nabla}\alpha=-(\zeta Mc^2/\alpha)\bm{\nabla}\alpha$.   Dicke
conjectured~\cite{dicke,dicke2} that
$\bm{\nabla}\alpha/\alpha\approx \alpha c^{-2}\bm{\nabla}\phi_N$. In
this view --  here designated the \textit{Coulomb model} regardless of
the law of $\alpha$ variation assumed --  the acceleration of $M$
comprises an anomalous fraction $\zeta\alpha$ which varies from
material to material (for example, the nuclear contribution to $\zeta$
ranges from 0.0016 for aluminum to 0.0041 for lead), thus engendering
a violation of the principle of universality of free fall of neutral
matter, a special case of the WEP.  Uzan~\cite{uzan} has given a
masterly review of this subject, as well as of the whole question of
$\alpha$ variability.

I made use of the Coulomb model for the anomalous force when drawing
conclusions from the general field-theoretic framework of $\alpha$
variability I formulated two decades ago~\cite{bek}.  The model has
also been uncritically adopted by most subsequent investigations in the
subject~\cite{livio,olive,dvali,MBS,SBM,BSM1,BSM2,otoole, Bmota}; a
refreshing exception is Landau et al~\cite{landau}.  Within the
general framework the Coulomb model predicts that the anomalous
acceleration is a fraction
$\sim \zeta^2$ of the total one; this is within an order of magnitude
of the fraction
$\zeta\alpha$ implied by Dicke's conjecture (see Ref.~[\onlinecite{bek}]
and Sec.~\ref{sec:naive} below).  Similar results are in evidence in
other treatments.  They have given rise to the widespread belief that
$\alpha$ variability necessarily implies violations of the WEP detectable
by E\" otv\" os-Dicke-Braginsky (EDB) type experiments~\cite{uzan},
particularly if the claimed cosmological variability~\cite{webb} is
essentially correct.

But as shown below,  such sweeping conclusion is unwarranted; it all
depends on the structure of the underlying field theory.  Contrary to
intuition and my original supposition, in the general
framework~\cite{bek} Coulomb energy of matter is found to be
unimportant as a source of the scalar field responsible for spatial
$\alpha$ variability.  Spatial gradients of that field are necessarily
much smaller than has been generally appreciated heretofore.  This
finding also impinges upon cosmology as it leads to a modification of
the widely used cosmological equation for
$\alpha$ variability.

The paper is designed as follows.  In Sec.~\ref{sec:equations} I
recapitulate the general variable $\alpha$ framework and discuss its
relation to other theories of $\alpha$ variability.  I rederive the
equations of motion for scalar and electromagnetic fields and particles
within the framework, and remark that one is prevented from assuming ---
as widely done --- that the dependence of particle masses on the scalar
field, which enters both in the anomalous force on particles and in the
source of scalar field, comes from Coulomb energy.  Rather the
equations of the theory themselves, being nonlinear, determine the
nature of their sources.

This is indeed seen from the \textit{exact} solution --- presented in
Sec.~\ref{sec:spherical} --- for the fields of a pointlike charge held
initially at rest in a uniform external electric field.  The dependence
of the particle's mass on scalar field is
\textit{not} of the sort expected from the Coulomb energy model: the
anomalous force is cancelled by a correction to the usual electric
force.  Thus there is no indication that different types of particle
with like charge-to-mass ratio would move differently in the same
external fields, and so there is no basis for a violation of the WEP.  I
also show that within the framework a charged particle has a minimum
possible extension, and that gravitational corrections to the mentioned
results are small, even for the most compact charge.

In Sec.~\ref{sec:general} I generalize the said solution to the case of
many charges held fixed in space. I prove rigorously that at an initial
moment, the center-of-mass acceleration of a \textit{collection} of
charges (some may be zero) starting at rest depends exclusively on its
total mass and charge, and on the external electric field, but not on
the structure of the collection.  Again no violation of the WEP is in
evidence.

Sec.~\ref{sec:macroscopic} shifts the focus from the microscopic
description of individual charges to the macroscopic description in
terms of smoothed electromagnetic and scalar fields and their smoothed
sources.  The macroscopic scalar field is shown by a variety of
examples to be too small to effect WEP violations detectable in the
foreseeable future.  Ordinary matter contributes two terms of Coulomb
origin to the source of the scalar field.  Taken together uncritically
these would suggest that in the vicinity of a chunk of matter,
$\bm{\nabla}\alpha/\alpha\sim
\zeta c^{-2}\bm{\nabla}\phi_N$, of the same order as Dicke's
conjectured spatial $\alpha$ variability.  But when the mentioned exact
solution is taken into account, the two Coulomb terms are seen to
cancel each other.  The source retains only terms of higher order;
these are incapable of generating WEP violating signals observable in
the foreseeable future.

Casting about for other sources of $\alpha$ variability, I show in
Sec.~\ref{sec:magnetic} that contributions of spin and orbital magnetic
dipoles in ordinary matter to the sources of the scalar field are
likewise too weak to engender violations of the WEP at soon-to-be
observable levels.

In Sec.~\ref{sec:real} I show that the cancellation of the Coulomb
energy in the source of the scalar field makes baryonic magnetic energy
the dominant source of cosmological
$\alpha$ variability.  This, together with the significant value of the
fundamental scale of the theory permitted by WEP tests (as newly
understood), makes it possible to understand the observed cosmological
growth of $\alpha$ and the rate of it as reflecting standard properties
of cosmological baryonic matter; there is no need for this to postulate
dark matter with peculiar electromagnetic properties.
Sec.~\ref{caveats} summarizes the conclusions and caveats on them.

Below I shall employ the signature $\{-1,1,1,1\}$ and the convention
that Greek indices range from 0 to 3, while Latin indices take on
values from 1 to 3.  The time coordinate is denoted by $t$ or $x^0$.

\section{\label{sec:equations}Framework for $\alpha$ variability}

\subsection{\label{sec:modifying}Modifying electrodynamics}

The general field-theoretic framework for $\alpha$
variability~\cite{bek} is based on eight assumptions: (1) for constant
$\alpha$ the framework's electromagnetism reduces to Maxwell's with a
minimal coupling to charged matter, (2)
$\alpha$ dynamics comes from an action, (3) this as well as
electrodynamics' action are relativistic invariants, (4) the overall
action respects gauge invariance, (5) electromagnetism is causal and
(6) respects time reversal-invariance, (7) any length scale in the
theory is not smaller than Planck's length $\ell_P=(\hbar
G/c^3)^{1/2}\approx 1.616\times 10^{-33}$ cm, and (8) gravitation is
governed by the Einstein-Hilbert action.

The choice of units that makes $G$, $\hbar$ and $c$ constant shifts the
burden of variation onto the charges.  Simplest is the case where all
charges $e_i$ vary in unison: $e_i=e_{0i}\,
\epsilon(x^\mu)$, where $e_{0i}$ denotes the coupling
\textit{constant} of particle $i$ and  $\epsilon(x^\mu)$ is a
\textit{dimensionless} scalar field (scalar since charge is an
invariant in relativity).  There is arbitrariness in the definition of
$\epsilon$; one can multiply it by a constant and divide all $e_{0i}$
by the same constant without changing anything.  That is why one must
demand that the dynamics of
$\epsilon$ be invariant under global rescaling of this field
(charge-scale invariance).  The only possible form of the free action
for
$\epsilon$ is thus
\begin{equation}  S_\epsilon=-{\hbar c\over 2
l^{2}}\int\epsilon^{-2}\epsilon_{,\mu}\epsilon_,{}^{\mu}(-g)^ {1/2}d^4x,
\label{eaction}
\end{equation}  where $g$ denotes the determinant of the metric
$g_{\mu\nu}$, $l$ is a constant scale of length introduced for
dimensional reasons; by assumption $l$ cannot be smaller than  $\ell_P$.

By assumption (1) $\epsilon$ must enter into all electromagnetic
interaction terms in the matter action via the replacement $e_i
A_\mu\mapsto e_{0i}\, \epsilon A_\mu$, with
$A_\mu$  the usual electromagnetic potential.    Gauge invariance of
the matter action [assumption (4)] will then be preserved only if a
gauge transformation means
\begin{equation}
\epsilon A_\mu\mapsto \epsilon A_\mu+\lambda_{,\mu}
\label{gauge}
\end{equation} with $\lambda$ any scalar function of spacetime point.
In order for the electromagnetic action to be invariant under this
transformation as well as under
$\epsilon$ rescaling, it must take the form
\begin{equation} S_{\text{em}}=-{1\over 16\pi}\int F_{\mu\nu}
F^{\mu\nu}(-g)^ {1/2}d^4x,
\label{EMaction}
\end{equation} where $F_{\mu\nu}=\epsilon^{-1}[(\epsilon
A_{\nu})_{,\mu}-(\epsilon A_\mu)_{,\nu}]$ obviously stands for the
gauge and $\epsilon$-scale invariant electromagnetic field tensor.  One
can consider adding to the integrand of
$S_{\text{em}}$ a term such as $ F_{\mu\nu} {}^\ast F^{\mu\nu}$, where
${}^\ast F^{\mu\nu}$ denotes the dual of $ F_{\mu\nu}$.  In Maxwellian
electrodynamics such an addition is equivalent to a boundary term, and
classically irrelevant.  Here this is not true because of the appearance
of the factor $\epsilon^{-2}$ in the integrand,  but the extra term
must nevertheless be rejected because it violates time-reversal
invariance [assumption (6)].

\subsection{\label{sec:actions}Actions}

The appearance of $\epsilon$ in the electromagnetic interaction means
that the equation for $\epsilon$ will involve $A_\mu$.  This was found
to be a bit inconvenient in Ref.~[\onlinecite{bek}].  Thus I shall here
adopt the Sandvik-Barrow-Magueijo (SBM) procedure~\cite{SBM} of
replacing
$A_\mu$ by another 4-potential, $a_\mu\equiv \epsilon A_\mu$.  The
gauge transformation (\ref{gauge}) now turns into
\begin{equation} a_\mu\mapsto a_\mu+\lambda_{,\mu},
\label{newgauge}
\end{equation} so that it is suitable to think of
\begin{equation} f_{\mu\nu}\equiv  a_{\nu,\mu}- a_{\mu,\nu}
\label{Faraday}
\end{equation}  as the new electromagnetic field tensor; it will turn
out to be the physical field tensor (Sec.~\ref{sec:eq_of_motion}).
Likewise I express
$\epsilon$ everywhere  in terms of SBM's field
$\psi=\ln\epsilon$.

The total action thus becomes
$S=S_{(\psi)}+S_{(f)}+S_{\text{m}}+S_{\text{g}}$, where
$S_{\text{m}}$ is the matter action (including the electromagnetic
interaction)
\begin{eqnarray} S_{(\psi)}&=&-{1\over 8\pi\kappa^2}
\int\psi_{,\mu}\psi_,{}^{\mu}(-g)^ {1/2}d^4x
\label{Spsi}
\\ S_{(f)}&=&-{1\over 16\pi}\int e^{-2\psi} f_{\mu\nu}f^{\mu\nu}(-g)^
{1/2}d^4x
\label{SM}
\\ S_{\text{g}}&=&{c^4\over 16\pi G}\int R(-g)^{1/2}d^4x
\label{gaction}
\end{eqnarray} with
\begin{equation}
\kappa\equiv {l\over(4\pi\hbar c)^{1/2}}\approx 8.11\times 10^{-26}
{l\over \ell_P}\ {\text{erg}^{1/2}\over
\text{cm}^{1/2}}
\label{kappa}
\end{equation}

In the new form of $S$, $\psi$ enters not only in $S_{(f)}$ but also
appears in the particle masses in $S_{\text{m}}$, even  for elementary
particles.  This point is clear from the example of a fermion particle
coupled to electromagnetism.  If charges were truly constant, the
process of renormalization would introduce $e_i$ dependence in the
fermion (dressed) mass.  After the transition
\begin{equation} e_i A_\mu\mapsto e_{0i}\, \epsilon A_\mu
\label{couple}
\end{equation} the mass becomes a function of $\epsilon$.  Swapping
$a_\mu$ for $A_\mu$ eliminates $\epsilon$ from the electromagnetic
interaction, but leaves $\epsilon$ or $\psi$ dependence in the mass.
It is intractable to calculate such dependence by summing the quantum
corrections to all orders in perturbation theory.  However, the
framework offers a self-consistent way to compute $m(\psi)$ which is
investigated in Appendix~\ref{sec:m}.

\subsection{\label{sec:other}Relation to other theories}

As discussed in Ref.~[\onlinecite{bek}] and already foreseen by Dicke
(Ref.~[\onlinecite{dicke}], Appendix 4), the second form of the framework
describes \textit{constant} charges in the presence of varying vacuum
permittivity and permeability,
$e^{-2\psi}$ and $e^{2\psi}$, respectively. The framework's action
differs from that in Dicke's theory only in that here the
electromagnetic Lagrangian is coupled to an exponential of the scalar
field (as required by charge-scale invariance of the original form of
the framework) whereas Dicke made the coupling linear in the field.
There is a resemblance also to Jordan's theory of varying natural
constants~\cite{jordan,uzan}; apart from the coefficients of the
various terms, the action recapitulated here is the case
$\eta=0$ of Jordan's.

This is the place to mention the variable speed-of-light (VSL) theories
~\cite{moffat} which have also been touted as variable $\alpha$
theories~\cite{magueijo,MBS}.  Dicke's dictum~\cite{dicke} (c.f.
Sec.~\ref{sec:intro}) that only variation of a dimensionless constant
is operationally meaningful does not preclude the formulation of a theory
which promotes $e$, \textit{or} alternatively $c$, to the  status of
a dynamical field.  It only asserts that when $e$ and $c$ appear in the
same physical context, the two forms of the theory would be
experimentally indistinguishable~\cite{bek1,duff}.  This point has led
to loud controversy with some authors affirming that a VSL theory makes
different predictions for $\alpha$ variability than a variable
$e$ theory~\cite{magueijo,MBS,moffat1}, while others deny
it~\cite{duff}.

This disagreement is easily defused; $c$ appears in physical actions in
at least four contexts: in the electromagnetic field-to-matter
coupling,  cf. Eq.~(\ref{pointparticle}), in mass terms, in
Lagrangian prefactors, e.g. the
$\hbar c$ prefacing the Dirac field action, and in the Einstein-Hilbert
action~(\ref{gaction}).  Only for the first of these is $c$ variation
fully swappable for $e$'s in the context of variable
$\alpha$.   In fact, the variable factor in the $c$ featuring in the
electromagnetic field-to-matter coupling could be absorbed into the
electromagnetic potential at the sole cost of introducing a dynamical
factor in the electromagnetic field action (\ref{EMaction}), which
would then take a form reminiscent of $S_f$, Eq.~(\ref{SM}) of the
framework studied here (a variable $e$ theory).  However, a theory
where at least two $c$'s with different roles are promoted to different
dynamical fields can obviously make predictions different from those of
a theory where only $e$ varies.  Such, for example, is Magueijo's
covariant VLS~\cite{magueijo}.  In the case that all matter is just
electromagnetic field, this theory can be understood as one where the
$c$ of electromagnetism and that from the gravitational action are
promoted to different powers of the scalar field.   This theory, which
essentially coincides with the general case of
Jordan's~\cite{jordan,uzan}, would predict a different cosmological
$\alpha$ evolution than does the present framework~\cite{MBS}.  But
that is because the modified Maxwellian electrodynamics in Magueijo's
theory is supplemented by a Brans-Dicke style modification of gravity:
it is a theory of variable $\alpha$ \textit{and} variable gravitational
coupling.

\subsection{\label{sec:eq_of_motion}Equations of motion}

Just as in other theories, in the framework the simplest
$S_{\text{m}}$ is that describing a pointlike classical, possibly
charged, particle:
\begin{widetext}
\begin{equation} S_{\text{m}}=\int\Big[-mc\Big(-g_{\mu\nu}{dz^\mu\over
d\tau}{dz^\nu\over d\tau}\Big)^{1/2}+{e_0\over c}{dz^\mu\over d\tau}
a_\mu\Big]c\gamma^{-1}\delta^3\big[\mathbf{x}-\mathbf{z(\tau)}\big]d^4x
\label{pointparticle}
\end{equation}
\end{widetext} where
$z^\mu(\tau)=\{z^0(\tau),\mathbf{z}(\tau)\}$ is the world line of the
particle as function of proper time $\tau$, and
$\gamma=dz^0/d\tau$ is its Lorentz factor.  The vanishing of
$\delta S_{\text{m}}/\delta x^\mu$, when combined with the condition
$g_{\mu\nu}dz^\mu dz^\nu = -c^2d\tau^2$, gives the equation of motion
\begin{equation} {D (mu_\alpha)\over d\tau}=-{\partial mc^2\over
\partial\psi}\psi_{,\alpha}+{e_0\over c}f_{\alpha\beta}u^\beta
\label{eqnmotion}
\end{equation} or
\begin{equation} m {Du_\alpha\over d\tau}=-{\partial mc^2\over
\partial\psi}(\psi_{,\alpha}+ u_\alpha u^\beta\psi_{,\beta})+{e_0\over
c}f_{\alpha\beta}u^\beta
\label{Lorentz}
\end{equation} where $D/d\tau$ stands for the covariant derivative
along the velocity $u^\alpha\equiv dz^\alpha/d\tau$, itself subject to
$g_{\mu\nu}u^\mu u^\nu = -c^2$.  It is plain from this that
$f_{\alpha\beta}$ is the physical electromagnetic field.
Eq.~(\ref{Lorentz}) makes it clear that in general,  in addition to the
(suitably modified) Coulomb and Lorentz forces, an anomalous force
coming from the
$\psi$ dependence of mass acts on any test charge immersed in a
background electromagnetic field.

The vanishing of $\delta(S_{\text{m}}+S_{(f)})/\delta a_\mu$ gives the
electromagnetic field equations
\begin{eqnarray} (e^{-2\psi}f^{\mu\nu})_{;\nu}&=&{4\pi\over c} j^\mu
\label{Maxwell}
\\ j^\mu&\equiv& e_0 c u^\mu
{\delta^3\big[\mathbf{x}-\mathbf{z(\tau)}\big]\over \gamma
\surd -g}
\label{current}
\end{eqnarray} The appearance of $e^{-2\psi}$ in Eq.~(\ref{Maxwell})
confirms the interpretation of this factor as permittivity (or
reciprocal of the permeability) of the vacuum in the present version of
the theory. The conservation of the current $j^\mu$ follows directly
from Eq.~(\ref{Maxwell}); the conserved charge is the truly constant
$e_0$.  It is in this sense that electric charge is still conserved in
this ``variable charge framework''~\cite{bek}.

Finally the vanishing of
$\delta(S_{\text{m}}+S_{(\psi)}+S_{(f)} )/\delta\psi$ gives the
equation for
$\psi$,
\begin{equation}
\psi_{,\mu;}{}^\mu =4\pi\kappa^2{\partial mc^3\over\partial
\psi}{\delta^3\big[\mathbf{x}-\mathbf{z(\tau)}\big]\over
\gamma
\surd -g}-{\kappa^2\over 2}e^{-2\psi}f_{\mu\nu}f^{\mu\nu}
\label{wave_equation}
\end{equation} while that of
$\delta(S_{\text{m}}+S_{(\psi)}+S_{(f)}+S_{\text{g}})/\delta
g^{\mu\nu}$ gives the gravitational field equations.

To fully specify the equation of motion of charged particles,
Eq.~(\ref{Lorentz}), one must specify $m(\psi)$.  As mentioned in
Sec.~\ref{sec:intro}, it has been customary to assume that the
spacetime dependence of mass reflects the electromagnetic contribution
to it, here proportional to $\epsilon^2$ or
$e^{2\psi}$.  Thus for ordinary matter (for which the Coulomb energy
far surpasses magnetic energy), it has been customarily assumed that a
fixed fraction $\zeta$ of $mc^2$ is Coulombic, so that $\partial
m/\partial\psi=2\zeta
mc^2$~\cite{bek,livio,olive,dvali,MBS,SBM,BSM1,BSM2,otoole, Bmota}.
Because $\zeta$ is expected to vary from object to object, violation of
the WEP would seem to be inevitable.  However --- and this is one of the
main points of the present paper --- the mentioned prescription for
$m(\psi)$ is the wrong one for the framework defined in
Sec.~\ref{sec:equations}.  The point is that the field equations
(\ref{Maxwell})--(\ref{wave_equation}) are nonlinear, e.g. the
permittivity $e^{-2\psi}$ is determined by the electromagnetic field
strength through Eq.~(\ref{wave_equation}).  It is well known that the
equations of motion of the sources of nonlinear field equations, e.g.
Einstein's equations, cannot be freely prescribed.  Rather, their
nature is specified by the field equations themselves.  Here, too, one
must let the field equations specify the nature of $m(\psi)$ which
defines the explicit form of the source's equation of motion,
Eq.~(\ref{Lorentz}).

\section{\label{sec:spherical}Isolated point electric charge}

\subsection{\label{sec:basic}The charge's fields}

An important step in clearing up the status of the WEP in the framework
is the study of the motion of a  pointlike charge in a specified
external electric field.  The  theory's nonlinearity does not permit
one to ignore the charge's own fields in setting up its equation of
motion; the task would be easier in Maxwellian electrodynamics where
fields can be superposed.    For now I neglect the curvature of
spacetime; as shown in Sec.~\ref{sec:justification}, this is entirely
justified for elementary particles and small collections of them.  I
also assume that initially the charge is held in place by some
unspecified force, and then released, so that its fields are static to
start with.

Setting $\gamma=1$ and $\mathbf{x(\tau)}=0$ in
Eqs.~(\ref{Maxwell})--(\ref{wave_equation}) one can look for a static
solution with vanishing magnetic field
$(f^{ij}=0)$.  With the notation
$c^{-1}\mathbf{E}\equiv \{f^{01},f^{02},f^{03}\}$ one gets
\begin{eqnarray}
\bm{\nabla}\cdot\left(e^{-2\psi}\mathbf{E}\right)&=&4\pi
e_0\delta^3(\mathbf{x})
\label{Phieqn}
\\
\bm{\nabla}^2\psi&=&4\pi\kappa^2\Big[{\partial mc^2\over\partial
\psi}\delta^3(\mathbf{x})+{1\over 4\pi}e^{-2\psi}\mathbf{E}^2\Big]
\label{psieqn}
\end{eqnarray} By Eq.~(\ref{Faraday}) and the obvious condition that
$a_\mu$ may be taken time independent, $\mathbf{E}$ must be a gradient:
$\mathbf{E}=\bm{\nabla}\Upsilon$.

The generic solution of Eq.~(\ref{Phieqn}) is
\begin{equation}
e^{-2\psi}\bm{\nabla}\Upsilon=-\bm{\nabla}\Phi+\mathbf{b};\quad
\mathbf{\bm{\nabla}\cdot b}=0,
\label{ansatz1}
\end{equation}  where
\begin{equation}
\Phi=e_0/r
\label{coulomb}
\end{equation}  is the usual Coulomb potential of the charge
$e_0$ in spherical polar coordinates
$\{r,\vartheta,\varphi\}$.  A particular solution is obtained when
$\mathbf{b}$ equals some
$\textit{constant}$ vector
$\bm{\mathfrak{E}}$.  Since $\psi$ should asymptote to a constant at
infinity, $\bm{\mathfrak{E}}$ is obviously the  applied external
electric field up to a positive proportionality constant.   The curl of
Eq.~(\ref{ansatz1}) gives
$\bm{\nabla}\psi \times\bm{\nabla}\Upsilon=0$ which shows that
$\psi$ and $\Upsilon$ are functions of each other.  By
Eq.~(\ref{ansatz1}) both can be taken as functions of only the
potential $V\equiv \Phi-\bm{\mathfrak{E}}\cdot\mathbf{x}$.  Thus
\begin{equation}
\mathbf{E}=-e^{2\psi}\bm{\nabla} V
\label{elecfield}
\end{equation} with $\psi=\psi(V)$.   So far it is clear that
Eq.~(\ref{elecfield}) is the unique solution when
$\mathbf{b}=\bm{\mathfrak{E}}$.   But as argued in
Sec.~\ref{sec:derive}, the uniqueness survives when
$\mathbf{b}$ is only known to asymptote to
$\bm{\mathfrak{E}}$.

In view of Eq.~(\ref{elecfield}),
\begin{equation}
\nabla^2\psi=\psi'\nabla^2 V +\psi'' e^{-4\psi}\mathbf{E}^2
\label{nabla}
\end{equation} ($'\equiv d/d V$, etc).  Since
$\nabla^2 V=-4\pi e_0\delta^3(\mathbf{x})$, the first (second) term on
the r.h.s. of Eq.~(\ref{nabla}) must match the first (second) term on
the r.h.s. of Eq.~(\ref{psieqn}).  The second condition, namely,
\begin{equation}
\psi'' = \kappa^2e^{2\psi}
\label{equation}
\end{equation} can be integrated after multiplication by
$\psi'$ to give
\begin{equation}
\psi'{}^2= \kappa^2(e^{2\psi}+\varpi)
\label{psi'}
\end{equation} with $\varpi$ a dimensionless  constant of integration.
Appendix
\ref{sec:solutions} obtains all solutions of this equation.

Identifying the pointlike source terms of Eqs.~(\ref{psieqn}) and
(\ref{nabla}) gives
\begin{equation}
\kappa^2(\partial mc^2/\partial\psi)=-e_0\psi' \quad
\text{at}\quad \mathbf{x}=0.
\label{dm}
\end{equation} This condition makes the two source terms equivalent
even when the $\delta$ function is somewhat smeared (I shall show below
that in this theory there is a lower bound to the radius of any charge,
so this smearing must occur).  As evident from the appearance in the
r.h.s. of a term linear in $e_0$, the dependence arising from this
equation is not automatically of the form
$m=m_0+\text{const.}\times e_0{}^2e^{2\psi}$ commonly adopted (Coulomb
model).

\subsection{\label{sec:which}Physical choice of $\varpi$ and WEP}

To find out about $\varpi$ I now work out the total force of electric
origin acting on the charge $ e_0{}$.  I make no attempt to separate
out the self field.   Immediately after the charge is released
Eqs.~(\ref{Lorentz}) and (\ref{elecfield}) give
\begin{eqnarray} md\mathbf{v}/dt&=&-(\partial
mc^2/\partial\psi)\bm{\nabla}\psi+ e_0 \mathbf{E}
\nonumber
\\ &=&-[(\partial
 mc^2/\partial\psi)\psi'+ e_0e^{2\psi}]\bm{\nabla} V.
\label{force0}
\end{eqnarray} It is to be stressed that the force on the r.h.s. here
comprises both the (modified) Coulomb force and the anomalous force.
Substituting $(\partial  mc^2/\partial\psi)$ from  Eq.~(\ref{dm}) and
$\psi'{}^2$ from Eq.~(\ref{psi'}) gives, after a cancellation, that
\begin{equation}
 md\mathbf{v}/dt=\varpi  e_0\bm{\nabla} V
\label{force'}
\end{equation}

The choice $\varpi=0$ in the solution corresponding to a charge $e_0$
is physically untenable.  It would mean that in the presence of an
arbitrary external field
$\bm{\mathfrak{E}}$, $e_0$ experiences no force whatsoever.  This is
contrary to all experience, and the theory developed here is supposed
to describe the real world. It must be, then, that $\varpi\neq 0$.  For
$\varpi\neq 0$ the force on the r.h.s. includes the usual self-force
term proportional to
$\bm{\nabla}\Phi$ which is always dropped in Maxwellian
electrodynamics; I do so too.  Thus
\begin{equation}
 md\mathbf{v}/dt=-\varpi  e_0\bm{\mathfrak{E}}
\label{force''}
\end{equation}

It would be hasty to reject outright all the  cases with
$\varpi\neq -1$ on the grounds that Eq.~(\ref{force''}) would  then have
the wrong form; after all $\bm{\mathfrak{E}}$ is the external electric
field only up to a positive proportionality constant.   However, with
$\varpi>0$ the net force acting on the particle \textit{would} be
opposite the accepted one.  One cannot remove this problem by assuming
that the ``passive'' charge here which senses the applied field is
opposite in sign to the ``active'' charge $e_0$ which is the source of
the electric field in Eq.~(\ref{Phieqn}).  For if this were true here,
it would be true in other cases too.   If
$\bm{\mathfrak{E}}$ is then interpreted as representing the  field of a
distant charge (up to a positive multiplier), it is immediately
apparent that charges of one sign would
\textit{attract}.  I thus conclude that physically
$\varpi$ must be negative.

In Appendix \ref{sec:solutions} all solutions of Eq.~(\ref{psi'}) are
found; that for negative $\varpi$ is ($\chi$ is an integration
constant):
\begin{equation} e^\psi=|\varpi|^{1/2}\sec(|\varpi|^{1/2}
\kappa V+\chi)
\label{soln}
\end{equation}      This solution is unique (in the physical sense) by
the one-to-one correspondence between solutions of Eq.~(\ref{psi'}) for
one charge and the solutions of the corresponding equation for many
charges (see Sec.~\ref{sec:derive}).  The latter are certified as a
complete set of solutions by the duality argument to be set forth in
Sec.~\ref{sec:monopole}, and the particular solution with $\varpi<0$
corresponds to Eq.~(\ref{soln}).

Let us interpret $e_p\equiv |\varpi|^{1/2}e_0$ as the physical charge
and $\bm{\mathfrak{E}}_p\equiv |\varpi|^{1/2}
\bm{\mathfrak{E}}$ as the physical external field.  This brings
Eq.~(\ref{force''}) to precisely the everyday form of the Newtonian
equation of motion in an external field;
$\varpi$ disappears from the equation of motion.

Eq.~(\ref{soln}) now takes the form
\begin{equation} e^\psi=|\varpi|^{1/2}\sec(\kappa V_p+\chi).
\label{soln'}
\end{equation} where $V_p\equiv
\Phi_p-\bm{\mathfrak{E}}_p\cdot\mathbf{x}$, and
$\Phi_p$  is built just as $\Phi$ in Eq.~(\ref{coulomb}) but from the
physical charge.  It would seem that $\varpi$ still appears after the
reinterpretation.  However, this is just an illusion.  Consider the
energy density of the electric field,
$\mathcal{E}=(8\pi)^{-1}e^{-2\psi}\mathbf{E}^2$ according to
Eq.~(\ref{T_Maxwell}).  This  can be recast in terms of  physical
quantities in such a way that $\varpi$ does not appear explicitly:
\begin{equation}
\mathcal{E}=(8\pi)^{-1}\sec^2(\kappa
V_p+\chi)\cdot(\bm{\mathfrak{E}}_p-\bm{\nabla}\Phi_p)^2
\end{equation} The same can be said about the purely Coulomb force on
the test charge, $\bm{\mathfrak{F}}= e_0\mathbf{E}$, which takes the
form
\begin{equation}
\bm{\mathfrak{F}}= e_p\sec^2(\kappa
V_p+\chi)\cdot(\bm{\mathfrak{E}}_p-\bm{\nabla}\Phi_p).
\end{equation} Again $\varpi$ has disappeared.  All this means that
setting $\varpi=-1$ (identifying $e_0$ with the physical charge) is a
matter of convention, and does not entail any assumptions about the
physics.  I thus take $\varpi=-1$ and drop the subscript $p$ henceforth.

The equation of motion~(\ref{force''}) for the particle in an external
field (with $\varpi=-1$) already shows that the WEP is not violated ---
at least not in an obvious way --- in the theory in question.  The
putative anomalous force from the
$\psi$ dependence of the mass is compensated for by the modification of
the Coulomb force effected by the $\psi$ field [the $e^{2\psi}$ factor
in Eq.~(\ref{force0})] in such a way that, at least for a
quasistationary charged particle, the force acting on the charge is
\textit{independent} of the fraction of its rest mass which is of
Coulomb provenance.  One finds no indication here for a violation of
the WEP.

A brief remark about neutral particles is in order here.  In that case
Eqs.~(\ref{Phieqn})-(\ref{psieqn}) reduce to a Poisson equation for
$\psi$ with a pointlike source.  The formal solution is $\psi=C/r$ with
$C$ a constant proportional to $\partial m/\partial \psi$ of the
source.  Dvali and Zaldarriaga~\cite{dz,dvali} suggest that $C$ is
generically nonzero with the consequence that long-range anomalous
forces exist even between neutral particles.  For the neutron they
infer this from the $\alpha$ dependence of the neutron mass coming from
virtual photon exchange between its constituent quarks~\cite{dvali} as
calculated perturbatively to low order.  As discussed in
Sec.~\ref{sec:derive}, the framework suggests rather that $C=0$, at
least when some charged particles are also present.

\subsection{\label{sec:size}Minimal size of an isolated charge}

Henceforth I set $\bm{\mathfrak{E}}=0$.  With $\varpi=-1$,
Eq.~(\ref{soln'}) gives
\begin{equation} e^\psi=\sec(\kappa\Phi+\chi).
\label{soln''}
\end{equation} Apart from the inclusion of the ``phase'' $\chi$, this
is the original solution presented in Ref.~[\onlinecite{bek}], where it
was pretty much gotten by guessing.  Redefining $\chi$ is equivalent to
shifting the zero of $\Phi$; this is certainly without physical
significance here as in Maxwellian electrodynamics.  As already clear
from Eq.~(\ref{coulomb}), I here adhere to the convention that
$\Phi$ vanishes at infinity.  The value of $\chi$ is thus fixed by the
asymptotic value of $e^\psi$, which coincides with the instantaneous
cosmological value of
$e^\psi$ in the appropriate model of the universe.

According to Sec.~\ref{sec:equations},  it is permissible to multiply
all $e_{0i}$ by a common positive constant while simultaneously
dividing $\epsilon$ by it.  I exploit this freedom to set the
cosmological value of $\epsilon$ to unity at the present epoch.  After
this is done one can define $\psi$ as the logarithm of $\epsilon$ and
pass to the second form of the theory.  From the
solution~(\ref{soln''}) at the mentioned epoch one then finds that
$\chi=0$ (this was also the choice of Ref.~[\onlinecite{bek}]).  Although
$\chi$ evolves cosmologically, it is evidently possible to set $\chi=0$
at any one cosmological epoch.  Thus results we shall obtain below
that depend on having $\chi=0$ are valid at any single epoch.  It is
only when interests centers on comparing physics at two separate epochs
that one can no longer do away with $\chi$.

By Eq.~(\ref{soln''}) $e^\psi$ can diverge and then turn negative when
$|\Phi|\geq \pi^{3/2}(\hbar c)^{1/2}l^{-1}$.  Because the permittivity
of the vacuum cannot be negative, this must mean that $|\Phi|$ can never
reach such values, i.e., that the particle with charge $e_0$ is spread
over a sufficiently large radius $R$ to prevent this.  By
Eq.~(\ref{coulomb}) this condition is
\begin{equation} R>\pi^{-3/2}(e_0{}^2 /\hbar c)^{1/2} (l/\ell_P)\ell_P.
\label{R}
\end{equation}

Because asymptotically $e^\psi\rightarrow 1$  at the present epoch, one
can take $(e_0{}^2 /\hbar c)^{1/2}=(137)^{-1/2}\sim 0.1$ for
$e_0$ the elementary charge.  And by the framework's assumption (7),
$l>\ell_P$; in fact I shall show in Sec.~\ref{sec:real} that if the
alleged cosmological $\alpha$ variability is to find explanation in this
framework, $l$ must be an order of magnitude above
$\ell_P$ (see also Refs.~[\onlinecite{bek,SBM,BSM1,MBS}]).  Thus the
\textit{lower bound} on the radius of any charge is at least a Planck
length.  Composite particles, e.g. the proton, easily satisfy
Eq.~(\ref{R}).  For leptons and quarks which are regarded as pointlike,
quantum gravitational effects must intervene at radii of a few Planck
lengths and modify the above classical considerations.  But it is
noteworthy that our formal lower bound on $R$ is not at variance with
the widespread belief that no elementary particle can be smaller than
$\ell_P$, the scale at which spacetime can no longer be regarded as a
continuum.

\subsection{\label{sec:justification}Why neglect spacetime curvature ?}

The neglect of spacetime curvature in all the preceding calculations
may be justified when the particle in question is either elementary or
made up of a not excessive number of elementary particles.
Appendix~\ref{sec:corrections} shows that corrections to the usual
exterior Reissner-Nordstr\" om metric (of general relativity) belonging
to a particle with charge $e_0$ become important in our theory only when
$|\Phi|$ is no longer small compared to $\pi^{3/2}(\hbar
c)^{1/2}l^{-1}$.  By Eq.~(\ref{R})
$|\Phi|$ gets that big only as $r$ approaches $R$.  Thus, if all one
wants is to investigate the source's exterior, one can, with good
accuracy, employ the Reissner-Nordstr\" om metric,
\begin{eqnarray} ds^2&&=-e^{-\lambda} dt^2+e^{\lambda} dr^2
+r^2(d\theta^2+\sin^2\theta d\varphi^2)
\label{metric}
\\ e^{-\lambda}&&\equiv 1-2Gmc^{_2} r^{-1}+Ge_0{}^2 c^{-4} r^{-2},
\end{eqnarray} where  $m$ is the source's mass.

This metric begins to departs seriously ($5\%$ level) from Minkowski's
at a radius $r\sim\max(5G^{1/2}e_0 c^{-2}, 40Gm c^{-2})$.  Now for the
known charged  elementary particles or small agglomerates of them,
$m\ll G^{-1/2}e_0$ (in fact
$G^{-1/2}e_0/m$ is of order $10^{25}$ for the electron and
$10^{20}$ for a nucleus or atom).  Thus significant departures from
flatness are only found at  $r < 5G^{1/2}e_0 c^{-2}=5(e_0{}^2/\hbar
c)^{1/2}\ell_P$, i.e., at Planck scale where the whole classical
description is already irrelevant.  And as pointed out in
Sec.~\ref{sec:size}, pointlike particles in this framework cannot be
smaller than this.  Thus, description of the exterior of a pointlike
elementary object (or a small collection of such) can well afford to
ignore spacetime curvature.

\section{Multiple electric charges}
\label{sec:general}

\subsection{\label{sec:derive}The solution}

The results above may be generalized to a collection of $N$ charges
$e_{0i}$ initially clamped at positions  $\textbf{z}_i$, $i=1, 2,
\cdots, N$.  Again with neglect of gravity
Eqs.~(\ref{Maxwell}--(\ref{wave_equation}) reduce to
\begin{eqnarray}
\bm{\nabla}\cdot\left(e^{-2\psi}\mathbf{E}\right)=4\pi
\sum_i e_{0i}\delta^3(\mathbf{x}-\mathbf{z}_i)
\label{Phieqn'}
\\
\nabla^2\psi=4\pi\kappa^2\left[\sum_i {\partial m_i
c^2\over\partial\psi}\delta^3(\mathbf{x}-\mathbf{z}_i)+{1\over
4\pi}e^{-2\psi}\mathbf{E}^2\right]
\label{psieqn'}
\end{eqnarray} As in Sec.~\ref{sec:spherical}, I take
$\mathbf{E}=\bm{\nabla}\Upsilon$.

I define anew
\begin{equation}
\Phi(\mathbf{x})=\sum_i {e_{0i}\over |\mathbf{x}-\mathbf{z}_i|};
\label{Phi'}
\end{equation} this is the standard Coulomb potential due to all the
charges.  Recalling that
\begin{equation}
\nabla^2\Phi=-4\pi\sum_i e_{0i}\delta^3(\mathbf{x}-\mathbf{z}_i),
\label{Poisson}
\end{equation} it is now easy to check that
\begin{equation}
\mathbf{E}=-e^{2\psi}\bm{\nabla}\Phi
\label{field}
\end{equation} together with any of the choices ($\chi$ is a constant)
\begin{equation} e^\psi=\left\{ \matrix{
\pm(\kappa \Phi+\chi)^{-1}; & \varpi=0 \cr
\\
\pm \surd\varpi\,\text{csch}( \surd\varpi\,\kappa
\Phi+\chi); &
\varpi>0 \cr
\\
 \surd|\varpi|\,\sec(\surd|\varpi|\,\kappa \Phi+\chi); &
\varpi<0 \cr },
\right.
\label{choices}
\end{equation}  constitute solutions of
Eqs.~(\ref{Phieqn'})-(\ref{psieqn'})  provided the analog of
Eq.~(\ref{dm}) is satisfied for every one of the charges, to wit
\begin{equation}
\kappa^2(\partial m_i c^2/\partial \psi)=-e_{0i}\psi' \quad
\textrm{at }\quad
\mathbf{x}=\mathbf{z}_i.
\label{dmi}
\end{equation} I shall exploit this last equation in
Appendix~\ref{sec:m} to elucidate the spatial variation of the
$m_i$.

It turns out that Eqs.~(\ref{field}) and (\ref{choices}) comprise all the
static solutions for multiple charges.  Although it is hard to prove
this directly, results in Sec.~\ref{sec:monopole} enable this be
established immediately by duality arguments.  The one-to-one
correspondence between the branches of $\psi$ in Eq.~(\ref{choices})
and those for the one-particle solution (\ref{summary}) then
establishes that in Sec.~\ref{sec:which}, Eq.~(\ref{elecfield}) is
indeed the most general form for $\mathbf{E}$ possible even if
$\mathbf{b}$ is not assumed constant except asymptotically;
generalizations of Eq.~(\ref{elecfield}) would necessarily affect
$\psi$ thorough Eq.~(\ref{psieqn}).

Since the physical multiparticle solution should include the physical
single particle solution as a special case, I put
$\varpi=-1$ here as in Sec.~\ref{sec:which}.   I also set
$\chi=0$ on the basis of the argument of Sec.~\ref{sec:size}.  That
this choice of parameters is consistent is made clear by the following
argument.  Suppose $\Phi_{i>1}$, the part of
$\Phi$ coming from particles $i=2, 3, \cdots\,, N$, is approximated by
the first two terms of a Taylor series about
$\mathbf{z}_1$.  Denoting $\kappa \Phi_{i>1}(\mathbf{z}_1)$ by
$\chi_1$ and $\bm{\nabla}\Phi_{i>1}$ at
$\mathbf{z}_1$ by $-\bm{\mathfrak{E}}$, the last of
Eqs.~(\ref{choices}) gives in the vicinity of charge $i=1$
\begin{equation}  
e^\psi\approx\sec\left\{\kappa\big[
\Phi_{i=1}-\bm{\mathfrak{E}}\cdot(\mathbf{x}-\mathbf{z}_1)\big]
+\chi_1\right\}.
\label{inducechi}
\end{equation}  This is of the same form as Eq.~(\ref{soln'}) for the
single charge solution.  Although the phase $\chi$ of the whole charge
complex has been set to zero, each individual charge has an associated
phase, $\chi_i$, induced by its neighbors.  The result
(\ref{inducechi}) makes it clear again that the one-charge solution
Eq.~(\ref{soln'}) is physically unique, as mentioned earlier.

With the above choices the full multiparticle solution is
\begin{eqnarray} e^\psi&=&\sec(\kappa\Phi)
\label{newpsi}
\\
\mathbf{E}&=&-e^{2\psi}\bm{\nabla}\Phi=-\kappa^{-1}\bm{\nabla}\tan(\kappa
\Phi)
\label{newelec}
\end{eqnarray} This result serves to clear up the question
(Sec.~\ref{sec:which}) about the value of $\partial m/\partial\psi$ for
a \textit{neutral particle}.  Because neutrinos and neutral mesons are
scarce in laboratory matter, the particle of most interest is the
neutron.  It is composite, but I avoid any discussion of its structure
and extension, and treat it as pointlike object as it would indeed
appear to low energy probes.   In this spirit one can include a neutron
as particle $n$ in the collection just discussed by formally taking
$e_{0n}\rightarrow 0$.  It is easy to verify that this limit is a
solution of the equations.  It follows from
Eqs.~(\ref{newpsi})--(\ref{newelec}) that $\psi$ is regular at the
neutral's position, and from Eq.~(\ref{dmi}) that
$\partial m_n/\partial\psi =0$.  More generally, for a
\textit{neutral} pointlike particle, $m$ is $\psi$ (and spacetime)
independent.  It remains a task for the future to reconcile this
conclusion with the dependence of the neutron mass on $\alpha$
according to perturbative calculations within QCD~\cite{dvali}.

\subsection{\label{sec:momentum}The momentum equation and WEP}

In our context the WEP would require that a composite possibly charged
body moves in {\it uniform\/} external electric and gravitational
fields with an acceleration which depends only on its total mass and
charge, but not on its detailed inner structure.  To put the framework
to the test in this respect, I imagine all $N$ charges in ``the world''
to be lumped into two clusters.  One, a spherical massive one, I put at
the origin and regard it as a single particle of mass $m_1$ and charge
$e_{01}$.  The other, denoted by $\mathcal{C}$, is made up of $N-1$
charges $e_{0i}$ with masses $m_{0i}$; $i=2, 3,
\dots\ N$.  By assuming the distance between the clusters is large
compared to both their extensions, it is possible to think of the
charge at the origin as pointlike, and
$\mathcal{C}$ as immersed in the uniform fields of the former.

I take the static solution for the fields,
Eqs.~(\ref{newpsi})--(\ref{newelec}), as part of the initial conditions
for the envisaged dynamical situation.  I restrict discussion to the
initial moment $t=0$ when all the charges are still at rest.   This
restriction is necessary here because the fully dynamical solution is
as yet unknown.   However, it should be clear that any violations of
the WEP would be expected to show up already at the initial moment
because they involve the acceleration.  The demonstration in
Sec.~\ref{sec:WEP} that no such violations occur, at least within the
approximation to be described presently, strongly suggests that the WEP
holds to great accuracy in the dynamical situation as well.

As a first step I calculate the \textit{rate} of change of total
momentum in the masses constituting $\mathcal{C}$.  According to
Eq.~(\ref{eqnmotion}) in the nonrelativistic approximation and with
neglect of gravity
\begin{equation}  {d\over dt}\sum_{i=2}^{N} m_i
\textbf{v}_i=\sum_{i=2}^{N}
\Big(-{\partial m_i c^2\over \partial\psi}\bm{\nabla}\psi+ e_{0i}
\mathbf{E}\Big)_{\mathbf{x}=\mathbf{z}_i}.
\end{equation} Substitution from Eqs.~(\ref{dmi}) and
(\ref{newpsi})--(\ref{newelec}) and use of the identity
$\sec^2x-\tan^2 x=1$ transforms this into
\begin{equation} {d\over dt}\sum_{i=2}^{N} m_i
\textbf{v}_i=-\sum_{i=2}^{N}
e_{0i}\bm{\nabla}\Phi|_{\mathbf{x}=\mathbf{z}_i}
\label{motion}
\end{equation}

Now near charge $i$, $\nabla\Phi$ is dominated by the self-field
$(\mathbf{x}-\mathbf{z}_i)/|\mathbf{x}-\mathbf{z}_i|^3$ which points
radially out from it.  As in Maxwellian electrodynamics, here one
may think of the force from this part as averaging out in the limit.
Left over from the force are the terms ($j\neq i$)
\begin{equation}
\sum_{i=2}^{N}\sum_{j=1}^N
{e_{0i}e_{0j}(\mathbf{z}_i-\mathbf{z}_j)\over
|\mathbf{z}_i-\mathbf{z}_j|^3}  =e_{01}\sum_{i=2}^{N}
{e_{0i}(\mathbf{z}_i-\mathbf{z}_1)\over |\mathbf{z}_i-\mathbf{z}_1|^3}
\nonumber
\end{equation} with the second form following cancellation of the
$j\neq 1$
 terms in pairs.  The assumed smallness of $\mathcal{C}$ as compared
with the distances $|\mathbf{z}_i-\mathbf{z}_1|$ justifies replacement
of every $\mathbf{z}_i$ on the r.h.s. by  the cluster's center-of-mass
position $\mathbf{Z}$ (an approximation  no different from the one
customarily made in Maxwellian systems):
\begin{equation} {d\over dt}\sum_{i=2}^N m_i
\mathbf{v}_i=Q{e_{01}(\mathbf{Z}-\mathbf{z}_1)\over
|\mathbf{Z}-\mathbf{z}_1|^3}
\label{newton0}
\end{equation} with $Q\equiv \sum_{i=2}^N e_{0i}$.   Thus the rate of
change of momentum of the particles in the cluster is controlled by the
formal Coulomb field of charge $i=1$, approximated as uniform at
$\mathcal{C}$, and by $\mathcal{C}$'s total charge.  Just as in
Maxwellian electrodynamics, here the cluster's internal structure
does not affect the rate of change of its total particle momentum.

Although this finding is consistent with the claim that the WEP is
satisfied here, it is no complete proof: the $m_i$ vary, making the
relation between rate of change of the total particle momentum and the
acceleration of the center-of-mass'  less clear than usual.

\subsection{\label{sec:WEP}Restricted microscopic proof of WEP}

To clarify the above point I look at the time component of
energy-momentum conservation for the whole system ($\mathcal{C}$
\textit{and} charge $i=1$).  If $T^{\mu\nu}$ includes the tensors for
the particles and the electromagnetic and scalar fields,
$T^{\mu\nu}{}_{;\nu}=0$ can be derived from the gravitational field
equations as usual.  Ignoring the gravitational field gives
\begin{equation}
\partial T^{00}/\partial t+\partial T^{0i}/\partial x^i =0.
\label{conserve}
\end{equation}   Multiplying the equation by $x^j$, integrating over a
large volume
$\mathcal{V}$ containing $\mathcal{C}$ but excluding  charge 1, and
integrating by parts gives
\begin{equation} {d\over dt}\int_\mathcal{V} T^{00}x^j
d^3x=\int_\mathcal{V} T^{0j} d^3x -\oint_{\partial\mathcal{V}}
x^jT^{0i} d^2S_i.
\end{equation} The integral on the l.h.s. is just the product of total
mass
$M=\int_\mathcal{V} T^{00}d^3x$ of the cluster (neglecting
contributions from the fields beyond the reaches of
$\mathcal{C}$) and the component $Z^j$ of $\mathbf{Z}$.  One further
time derivative gives
\begin{equation} {d^2\over dt^2}\int_\mathcal{V} T^{00}x^j d^3x={d\over
dt}\int_\mathcal{V} T^{0j} d^3x -{d\over dt}\oint_{\partial\mathcal{V}}
x^jT^{0i} d^2S_i.
\label{mom_cons}
\end{equation}

The spatio-temporal component $T^{0i}$ figuring in Eq.~(\ref{mom_cons})
receives three contributions.  The first,
$T_{(f)}{}^{0i}$ from the $f^{\mu\nu}$ field (see
Appendix~\ref{sec:corrections}), comprises products of
$f^{0i}$ with $f^{ij}$. The $f^{ij}$ evidently vanish at
$t=0$ (no motion, no magnetic field).   Further, because I assume the
static solution (\ref{newpsi})-(\ref{newelec}) for
$\psi$ and
$\mathbf{E}$ holds at $t=0$,  $f_{0i}=-c E^i$ is a gradient.  Then the
identity $\dot f_{ij}+f_{0i,j}+f_{j0,i}=0$ ($f_{\mu\nu}$ derives from a
4-potential) shows that initially
$\dot f_{ij}=0$.  Therefore, time differentiation anihilates both
integrals over $ T_{(f)}{}^{0i}$ in Eq.~(\ref{mom_cons}) at $t=0$ (that
$\dot\psi=0$ at $t=0$ will be shown presently).

The second contribution, $T_{(\psi)}{}^{0i}$ (again see
Appendix~\ref{sec:corrections}), comprises products of
$\psi^{,i}$ with
$\dot\psi$.  Now because the scalar equation (\ref{wave_equation}) is
of second order in time, one may require $\dot\psi=0$ at $t=0$; this is
consistent with this instant representing the end of a purely static
situation.    Further, comparison of Eqs.~(\ref{psieqn'}) and
(\ref{wave_equation}), the latter satisfied identically by solution
(\ref{newpsi})--(\ref{newelec}), shows that also
$\ddot\psi=0$ at $t=0$ for the assumed initial conditions.   Thus  time
differentiation of both integrals over
$T_{(\psi)}{}^{0i}$ in Eq.~(\ref{mom_cons}) fails to produce
nonvanishing contributions at $t=0$.

Accordingly, the time derivatives at $t=0$ in Eq.~(\ref{mom_cons}) of
both integrals over $ T{}^{0i}$ come solely from the particles in
$\mathcal{C}$.  Since the surface $\partial\mathcal{V}$ lies beyond
them, the r.h.s. of Eq.~(\ref{mom_cons}) comprises solely
${d\over dt}\int T_{\text{m}}{}^{0j}d^3x$.  The tensor in question is
derived from action (\ref{pointparticle}) by variation of
$g^{\mu\nu}$ and use of $g_{\mu\nu}u^\mu u^\nu = -c^2$:
\begin{equation} T_{\text{m}}{}^{\mu\nu}=\sum_i m_i c {dz_i^\mu\over
d\tau}{dz_i^\nu\over
d\tau}{\delta^3\big(\mathbf{x}-\mathbf{z}_i(\tau)\big)\over
\gamma\surd -g}
\end{equation} Neglecting the difference between $\tau$ and
$t$, replacing
$\surd -g$ by its Minkowski value $c$, and substituting in
Eq.~(\ref{mom_cons}) gives
\begin{equation} {d^2(M Z^j)\over dt^2} = {d\over dt}\sum_{i=2}^N m_i
\mathbf{v}_i.
\label{newton}
\end{equation}

I complete the derivation by showing that at $t=0$ $M$ can be taken out
from under the derivatives.  Integrating Eq.~(\ref{conserve}) over
$\mathcal{V}$ and using Gauss' theorem gives
\begin{equation}  {d\over dt}\int_\mathcal{V} T^{00} d^3x=
-\oint_{\partial\mathcal{V}} T^{0i} d^2S_i.
\end{equation}  Differentiating this result by $t$ gives
\begin{equation}  {d^2\over dt^2}\int_\mathcal{V} T^{00} d^3x=-{d\over
dt}
\oint_{\partial\mathcal{V}} T^{0i} d^2S_i.
\end{equation}  By precisely the same arguments given above, the r.h.s.
of these two equations vanish at $t=0$.  Since the integral in both
l.h.s. is  $\mathcal{C}$'s total mass $M$, this demonstrates that $\dot
M=\ddot M=0$ at
$t=0$.  Of course this does not mean that $M$ is a conserved quantity,
only that it behaves as $M(t=0)+\mathcal{O}(t^3)$ for short times.
Substituting this result in Eq.~(\ref{newton}) and taking
Eq.~(\ref{newton0}) into account gives at $t=0$
\begin{equation}
\ddot\mathbf{Z}= {Q\over M}{e_{01}(\mathbf{Z}-\mathbf{z}_1)\over
|\mathbf{Z}-\mathbf{z}_1|^3}.
\label{acc}
\end{equation}

Thus in harmony with the WEP, the acceleration of
$\mathcal{C}$'s center-of-mass  in the field of the distant charge
$e_{01}$ is fully determined by its mass $M$ and total charge $Q$, and
is insensitive to its structure (disposition of the member charges,
their charge to mass ratios, etc.)  To reach this result the mass $M$
of $\mathcal{C}$ had to be identified with the integral of $T^{00}$
taken over a
\textit{finite}, albeit large, region.  This is a necessity in any
situation when the system of interest is not the only one in the
universe; the same procedure would be required in any other field
theory.  Another limitation of the approach is that result (\ref{acc})
is rigorously valid only when all particles in
$\mathcal{C}$ are assumed to be at rest (thus the approach neglects
purely magnetic effects).  But intuitively the acceleration's
universality property should remain valid if all velocities are small
and spin magnetism is weak, as I indeed show in Secs.~\ref{sec:psi} and
\ref{sec:dipole}.

The results here concur with those reached by  Landau,  Sisterna and
Vucetich~\cite{landau} on the basis of the TH$\epsilon\mu$ formalism, a
nonrelativistic generic parametrized microscopic description of the
gravitational, particle and electromagnetic sectors of field theories.
The authors find a connection between the lack of overall charge
conservation and violations of the WEP.  Looking at the present
framework in the light of experimental constraints on charge
nonconservation, they are able to certify that WEP is respected to a
fractional accuracy (in the acceleration) $\sim 10^{-18}$, beyond the
projected sensitivity of WEP tests in the foreseeable future.  Since
charge conservation is actually exact in the framework [see the comments
after Eq.~(\ref{current})] the implication is actually stronger:  no
WEP violations are expected, at least nonrelativistically, even when
slow motion of charges is allowed.

There exist other possibilities for variability of $\alpha$ by way of a
scalar field which do not run counter to the EDB experiments.  Such is
the supersymmetric grand unified theory of Chacko, Grojean and
Perelstein~\cite{grojean}, according to which a late epoch cosmological
phase transition causes a jump in $\alpha$ while generating a vacuum
expectation value of the scalar field which makes it short ranged.
Consequently, although a charged particle's mass can be scalar field
dependent, this does not lead to long range anomalous forces which
would contradict the said experiments.

\section{Macroscopic description of electric structure and WEP}
\label{sec:macroscopic}

A virtue of the ``proof'' of the WEP in Sec~\ref{sec:WEP} is that it
does not require knowledge of the equations of motion for macroscopic
matter [as opposed to the microscopic level Eq.~(\ref{Lorentz})]. Its
shortcomings are that it establishes the validity of the WEP only for
brief time intervals after a putative quiescent situation, and only
when the magnetic structure of the systems involved is ignored.  Both
shortcomings will now be remedied. By averaging over microscopic
quantities to produce a macroscopic (or continuum) version of the
theory, I show in Sec.~\ref{sec:psi} that the WEP-breaking effects
pointed out in Refs.~[\onlinecite{dicke,dicke2,bek,livio,olive,dvali,MBS}]
are ruled out if one is willing to assume that a macroscopic chunk of
matter moves according to Eq.~(\ref{Lorentz}) with a natural definition of
$m(\psi)$ obtained by macroscopic averaging.  No short-times assumption
is then necessary.  And in Sec.~\ref{sec:dipole} magnetic dipole
structure of matter is incorporated into the arguments.

Secs.~\ref{sec:spherical} and\ref{sec:general} describe matter
microscopically, that is as a collection of pointlike particles $i$
with definite charges $e_{0i}$ (some of which may be zero) and masses
$m_i$ subject to the relation (\ref{dmi}).   One key assumption I make
on the way to the macroscopic description is that the many charges
solution  (\ref{newpsi})--(\ref{newelec}) remains valid, at least
approximately, when the particles move slowly.  I define the spatial
average $\bar\mathcal{Q}$ of a quantity $\mathcal{Q}$ as the integral
of $\mathcal{Q}$ over some macroscopic region of volume $\mathcal{V}$,
all divided by  $\mathcal{V}$.  In flat spacetime or in local Lorentz
frames (assumed to be large enough to encompass the said macroscopic
region), no ambiguity in this definition arises from issues of parallel
transport of vectors and the like.

Another assumption I make is that for a macroscopic test body
$\mathcal{T}$  of mass $\mu$ and charge $q$ moving on background
scalar, Newtonian and electric fields, $\bar\psi,\,
\bm{\nabla}\phi_N$ and $\bm{\mathfrak{E}}$ (all regarded as
approximately uniform), moves according to Eq.~(\ref{Lorentz})
[Eq.~(\ref{force0}) in the nonrelativistic case]:
\begin{equation}
\mu d\mathbf{v}/dt=-\mu\bm{\nabla}\phi_N-(\partial \mu
c^2/\partial\bar\psi)\bm{\nabla}\bar\psi+ q \bm{\mathfrak{E}}
\label{test}
\end{equation} The assumption would be a triviality but for the
stipulation that $ \mu(\bar\psi)$ is to be identified with the
macroscopic average of the total energy density in
$\mathcal{T}$ multiplied by its volume.  Evidently $q$ is to be
interpreted as the sum of charges in $\mathcal{T}$, while
$\bm{\mathfrak{E}}$ is the electric field determined from its sources
by Eq.~(\ref{Phieqn'}) and averaged over the volume of
$\mathcal{T}$.  The only quantity which requires special discussion is
$\bm{\nabla}\bar\psi$.

\subsection{The Coulomb model for $\bar\psi$}
\label{sec:naive}

In the language of the present framework, Dicke's argument~\cite{dicke}
amounts to assuming that the macroscopic field $\bar\psi$ is related to
its source
$\mathcal{S}$ via a macroscopic version of Eq.~(\ref{psieqn'}) with the
sum of  $(\partial m_i c^2/\partial \psi
)\delta^3(\mathbf{x}-\mathbf{z}_i)$ replaced by
$2\zeta \rho c^2$, where $\zeta$ is the typical fraction of the
source's mass density $\rho$ which is of Coulomb provenance.  This
comes from assuming that
$\rho\propto\alpha\propto e^{2\bar\psi}$.  Now the second term in the
square brackets in Eq.~(\ref{psieqn'}) is, according to
Appendix~\ref{sec:corrections}, exactly twice the density of electric
energy, just as the former contribution.  Hence in the Coulomb model
for the source, Eq.~(\ref{psieqn'}) takes the form
\begin{equation}
\nabla^2\bar\psi=16\pi\kappa^2\zeta\rho c^2=(l/\ell_P)^2(\zeta/\pi c^2)
4\pi G\rho,
\label{psi_phi}
\end{equation} A likeness of this equation and/or its cosmological version
reappears in subsequent
treatments~\cite{bek,livio,olive,MBS,SBM,BSM1,BSM2,otoole,Bmota}.

The similarity of Eq.~(\ref{psi_phi}) with Poisson's equation for the
Newtonian potential of $\mathcal{S}$, and the similarity between the
asymptotic boundary conditions on
$\phi_N$ and
$\bar\psi$, permits the identification
\begin{equation}
\bm{\nabla}\bar\psi = (l/\ell_P)^2(\zeta/\pi c^2)\bm{\nabla}\phi_N
\label{dicke_like}
\end{equation}  Actually, in a nonspherical configuration the curl of
some vector may be added to the r.h.s.; however, this ``correction''
must decay asymptotically as $1/r^3$ (cf. Ref.~\onlinecite{bekmil},
Appendix A) and so should become irrelevant some distance outside
$\mathcal{S}$.  Eq.~(\ref{dicke_like}) shows that
$\bm{\nabla}\alpha/\alpha= (l/\ell_P)^2(2\zeta/\pi
c^2)\bm{\nabla}\phi_N$.  With $l\sim \ell_P$ this result is very much
like Dicke's conjectured one~\cite{dicke} mentioned in
Sec.~\ref{sec:intro}, except for the replacement of $\alpha$ by $2\zeta$,
which for ordinary matter is of the same order as
$\alpha$.

How does this all bear on the WEP ?  Suppose $\mathcal{T}$  moves in
the vicinity of $\mathcal{S}$.  In the Coulomb model one replaces
$(\partial
\mu c^2/\partial\psi)$ in the equation of motion for
$\mathcal{T}$, Eq.~(\ref{test}), by
$2\zeta'\mu c^2$, where $\zeta'$ is now the Coulomb energy fraction of
$\mu c^2$. With the replacement (\ref{dicke_like}) this gives (in the
absence of other sources)
\begin{equation}
d\mathbf{v}/dt=-[1+(l/\ell_P)^2(2\zeta\zeta'/\pi)]\bm{\nabla}\phi_N +q
\bm{\mathfrak{E}},
\label{anomalous}
\end{equation} a version of which first appeared in
Ref.~[\onlinecite{bek}].  Now, as mentioned in Sec.~\ref{sec:intro}, for
ordinary matter $\zeta$ is of order a few times $10^{-3}$ and varies by
about $10^{-3}$ from material to material.  The latest EDB tests of WEP
find that
$d\mathbf{v}/dt$ is $\zeta'$ independent to fractional accuracy
$10^{-13}~$\cite{uzan}.  This is consistent with Eq.~(\ref{anomalous})
only if $l$ is  $10^{-3}\ell_P$ or smaller.  Because of this and
assumption (7) of the framework, I inclined in Ref.~[\onlinecite{bek}] to
the opinion that there is no $\alpha$ variability in nature (as would be
the case if
$l\equiv 0$, e.g. electrodynamics exactly Maxwellian).

This was also the conclusion of Livio and Stiavelli~\cite{livio} who
noted the difficulty in explaining the alleged cosmological $\alpha$
variability with
$l$ as small as $10^{-3}\ell_P$ and the accepted matter content of the
universe.  Olive and Pospelov~\cite{olive} also took cognizance of this
problem; to solve it they proposed that the cosmological dark matter is
much more strongly coupled to the
$\psi$ field than is ordinary matter.  This would have the effect of
counteracting the smallness of $l/\ell_P$ inferred from tests of the
WEP.  By contrast, Magueijo, Barrow and Sandvik~\cite{MBS} see no
immediate strong contradiction between the EDB experiments tests; they
infer
$(l/\ell_P)^2\approx 10^{-4}$ from the claimed cosmological
$\alpha$ variability by assuming that \textit{cosmological matter} has
$\zeta\approx 1$.  They further adopt, for ordinary matter, the very
low values
$\zeta\sim\zeta'\sim  10^{-4}$.  However, these run on the face of
simple estimates from nuclear Coulomb energy (see Sec.~\ref{sec:lab}). 
There is thus a tension between the claimed cosmological
$\alpha$ variability and the tests of the WEP.  In the next subsections
I show that the blame for the above \textit{impasse} lies squarely with
the misuse of the Coulomb model, there being no need to take
$l<\ell_P$.  When needed in what follows I assume  $l$ is above
$\ell_P$ and within an order of magnitude of it.

\subsection{\label{sec:psi}Coulomb energy and WEP violation}

For a quick orientation one can describe the detailed electric
structure of the source $\mathcal{S}$ of $\bar\psi$ with the static
electric solution, Eqs.  (\ref{Phi'}), (\ref{newpsi}--(\ref{newelec}).  One
is thus neglecting effects of internal motions and magnetic structure, both
usually minor complications for nonrelativistic sources of interest.

I assume that $\kappa|\Phi|\ll 1$ everywhere and check this on a
case-by-case basis.  Then Taylor expanding both sides of
Eq.~(\ref{newpsi}) and averaging as stipulated in the preamble of this
section gives
\begin{equation}
\bar\psi\approx {\scriptstyle 1\over\scriptstyle
2}\kappa^2\overline{\Phi^2}={\scriptstyle 1\over\scriptstyle
2}\kappa^2\bar
\Phi^2 + {\scriptstyle 1 \over\scriptstyle
2}\kappa^2\overline{(\Phi-\bar\Phi)^2}
\label{barpsi}
\end{equation} Note that matter with electric structure causes the
physical $\alpha$, which is proportional to
$e^{2\psi}\approx 1+2\psi$, to be slightly
\textit{larger} nearby than asymptotically (using the Coulomb model
Refs. \onlinecite{MBS,SBM} predict an effect with the opposite sign and
much larger magnitude).   Now
\textit{outside} a macroscopic source $\mathcal{S}$ the fluctuation
term here should be relatively small compared to
$\bar \Phi^2$, except in rather artificial situations where
$\bar\Phi$ very nearly vanishes (the net charge and some higher multipoles
are exactly zero).  The latter are not important in our context, so
henceforth I shall drop the fluctuation and write just $\Phi$ for
$\bar\Phi$.

\subsubsection{\label{sec:natural}Natural sources of gravity and
$\bar\psi$}

First I look at natural sources of $\bar\psi$, e.g. the Sun and Earth
in the EDB experiments.    Unless $\mathcal{S}$ is (almost) exactly
neutral, $\Phi$ is dominated by its monopole part.  For example, Earth
is known to bear a net charge at any time.  Thus for a quasispherical
source $\Phi\approx Q r^{-1}$ and $|\bm{\nabla}\bar\psi|\approx
\kappa^2 Q^2 r^{-3}$ at distance
$r=|\mathbf{r}|$ from the source's center.

There is a natural bound on $Q$ if the source, a natural large object
like Sun or Earth, is to be quiescent:
$|Q|<GMm_p/e_{0p}$, where $M$ is
$\mathcal{S}$'s mass, and $m_p$  and $e_{0p}$ the proton's mass and
charge, respectively.  For if $Q>GMm_p/e_{0p}$, the source's electric
field can drive away any free protons formed nearby by, say, cosmic ray
ionization of hydrogen even against the pull of gravity, while
$\mathcal{S}$ captures the electron and so decreases $Q$.  And if
$Q<-GMm_p/e_{0p}$,
$\mathcal{S}$  can certainly drive away the free electrons, capture the
protons and so decrease $|Q|$.  (In the above argument it is important
that $e^{2\psi}\approx 1$ by assumption, so
$\mathbf{E}\approx -\bm{\nabla}\Phi$.)

In view of the restriction on $Q$,
\begin{eqnarray}
\kappa|\Phi|&<& ({1\over 4\pi\alpha})^{1/2}\big({l\over
\ell_P}\big){\ell_P\over
\lambda_p} {GM\over c^2 r}
\nonumber
\\ |\bm{\nabla}\bar\psi|&<&{1\over 4\pi\alpha c^2}\big({l\over
\ell_P}\big)^2\big({\ell_P\over\lambda_p}\big)^2 {GM\over r^2}{GM\over
c^2 r}
\label{grad_Phi}
\end{eqnarray} where $\alpha=e_{0p}^2/\hbar c\approx 0.0073$ and
$\lambda_p=\hbar/m_p c\approx 2\times 10^{-14}$ cm is the proton's
Compton length.   Obviously $GM/c^2 r<1$ because the source is not a
black hole. It follows that
$\kappa|\Phi|<3\times 10^{-19}(l/\ell_P)\ll 1$, thus verifying the
initial assumption.   Further, $GM/r^2\approx |\bm{\nabla}\phi_N|$.
Thus
$|\bm{\nabla}\bar\psi|$ is a  fraction $2.2\times 10^{-37}\
\zeta^{-1}$ or smaller of the Coulomb model prediction
(\ref{dicke_like}).    In view of Eq.~(\ref{anomalous}), violations of
the WEP reflecting $\mathcal{S}$'s electric structure show up only at the
acceleration fractional level
$ 10^{-40}(l/\ell_P)^2$ or smaller.  (In light of the conclusions of
Secs.~\ref{sec:which} and \ref{sec:WEP}, the tiny residual WEP
violation here, and similar ones below, may well be artifacts of the
macroscopic averaging procedure.)  As mentioned already the extant
laboratory EDB experiments are sensitive only to acceleration
fractional differences at the level of $10^{-13}$.  The planned
satellite borne STEP experiments are projected to be sensitive only at
the level
$10^{-18}$.

What if $\mathcal{S}$ is nearly neutral, so that $\Phi$ is dominated by
its dipole moment $\mathbf{p}$: $\Phi\approx
\mathbf{p}\cdot
\mathbf{r}\ r^{-3}$.  In that case $|\mathbf{p}|$ should be on the
order of the radius $R$  of $\mathcal{S}$ times a typical separated
charge
$|Q_s|$.  The magnitudes of the charges
$\pm Q_s$, which pictorially reside on opposite polar caps of
$\mathcal{S}$, are restricted by the same inequality as the monopole charge
above.  Otherwise free electrons near the negative polar cap would
be driven away against gravity and conveyed by the dipole field to the
positive cap, thus helping to diminish $|\mathbf{p}|$.  Repeating the
above argument one finds  the expressions for $\kappa|\Phi|$ and
$|\bm{\nabla}\bar\psi|$ to be similar to those in Eqs.~(\ref{grad_Phi})
but each with an extra factor
$(R/r)^2<1$. Thus $\kappa|\Phi|\ll 1$ is still satisfied, and the
failure of the Coulomb model is accentuated.  Evidently higher
multipoles do not offer a way out of the conclusion that the Coulomb
model is very far off the mark.   Evidently laboratory and space tests
are far from sensitive to violations of the WEP coming from the
electric structure of the sources of gravity.

\subsubsection{\label{sec:lab}Laboratory  sources of gravity and
$\bar\psi$}

But what if the source $\mathcal{S}$ is not a gravitating body in the
dirty interplanetary environment, but rather a mundane body in a clean
laboratory where its charge does not immediately get neutralized ?
Would the Coulomb model apply then ?  An example might be furnished by
a lead sphere of radius
$R$.  Again Eq.~(\ref{barpsi}) tells us that if the sphere holds charge
$Q$, then outside it
$|\bm{\nabla}\bar\psi|\approx
\kappa^2 Q^2 r^{-3}<\kappa^2 (Q^2/R)r^{-2}$.  The factor
$Q^2/R$ is twice the formal Coulomb energy associated with the net
charge $Q$.  How big can $Q$ be here ? Unless the sphere is in an
evacuated cell (which procedure can buy us a few orders of magnitude in
the discouraging results below),  it can only be charged until its
surface electric field reaches the air breakdown level,
$\sim 3\times 10^4\, \text{V}/\text{cm}=1\times 10^2
\,\text{esu}$.   At that point the sphere is at $\Phi\sim 1\times 10^2\,
\text{esu}\ R$.  According to Eq.~(\ref{kappa}),
$\kappa\approx 8.11\times 10^{-26}\, \text{esu}$.  Thus  indeed
$\kappa|\Phi|\ll 1$, as assumed.  In said state the sphere holds  $\sim
1\times 10^2\, \text{esu}\  R^2$  of free charge, and thus twice its
macroscopic Coulomb energy is
$\sim 1\times 10^{4}\, \text{erg}/\text{cm}^3\, R^3$. Hence
\begin{equation} |\bm{\nabla}\bar\psi|<1\times 10^{4}\
\text{erg}/\text{cm}^3\ \kappa^2 R^3 r^{-2}
\label{true}
\end{equation}

By contrast, in the Coulomb model Eq.~(\ref{psi_phi}) would predict
\begin{equation} |\bm{\nabla}\bar\psi|=4\kappa^2[\zeta (4\pi R^3/3)
\rho  c^2]r^{-2},
\label{coulomb_model}
\end{equation}  with the square brackets recognizable as the
\textit{microscopic} level Coulomb energy.  Being lead
($\rho=1.13\times 10^1\, \text{g}/\text{cm}^3$ and mass number 207.2) the
sphere contains $\sim 1.38\times 10^{23}/\text{cm}^3 R^3$ Pb nuclei;
each contributes $\sim 795$ MeV of Coulomb energy~\cite{segre} for a
total of $\sim 1.75\times 10^{20}\
\text{erg}/\text{cm}^3\ R^3$.  Thus in the present example the true
$|\bm{\nabla}\bar\psi|$, Eq.~(\ref{true}), amounts to a fraction
$1.4\times 10^{-17}$ of the  Coulomb model prediction, independent of
$R$.   In this case also the Coulomb model yields a resounding
overestimate.  In view of  Eq.~(\ref{anomalous}), violations of the WEP
for test body motion subject only to the field of a macroscopically
charged laboratory sized object are at a fractional level
$10^{-23}(l/\ell_P)^2$.  But this fraction is actually suppressed to
$10^{-31}(l/\ell_P)^2$ because  $\mathcal{S}$'s gravitational field is
diluted by a factor $10^8$ by Earth's gravitational field (for this
calculation I take
$\mathcal{S}$'s mass to be $10^6\, \text{g}$).  Thus the WEP violation
considered here is well below the sensitivity of the laboratory EDB
experiments.  The planned STEP experiments are irrelevant in this
context because they do not include their own source of
$\bar\psi$.

The prospects for WEP violation improve substantially when
macroscopically charged sources in the laboratory are replaced by
macroscopically polarized ones in Earth orbit (an extension of the STEP
experiment which is planned to carry only test objects).  A good
example is a sphere of radius $R$ made of ferroelectric material.  In a
ferroelectic there is spontaneous alignment of the molecular electric
dipoles, so that a macroscopic polarization vector
$\mathbf{P}$ appears.  Thus in Eq.~(\ref{barpsi}) $\Phi\approx (4\pi
R^3/3)\mathbf{P}\cdot\mathbf{r}\ r^{-3}$.  A good estimate of the
maximum $|\mathbf{P}|$ is one elementary dipole
$e_{0p}d$ per molecular volume $d^3$.  With  $d\sim 10^{-7}$ cm this
works out to
$|\mathbf{P}|< 4.8\times 10^{4}\, \text{esu}$.  Indeed PbTiO$_3$, a
ferroelectric with one of the largest measured polarizations, shows
$|\mathbf{P}| \sim 1.5\times 10^{5}\,
\text{esu}$~\cite{kittel}. Given that $r>R$, one finds
$\kappa|\Phi|<(4\pi)^{-1/2}\alpha^{1/2} lR d^{-2}$.   With
$R\sim 10^2$ cm this bound is below $10^{-18}\, (l/\ell_P)$, small as
originally assumed.

Using Eq.~(\ref{barpsi}) one estimates
\begin{equation} |\bm{\nabla}\bar\psi|\approx 4\kappa^2 ({R/r})^3[(4\pi
R^3/3)
\mathbf{P}^2]r^{-2}
\label{dipol}
\end{equation} Eq.~(\ref{dipol}) exhibits an energy
$\mathbf{P}^2 d^3< (e_{0p} d)^2/d^{3}$ (perhaps tens of eV per
molecule) where the Coulomb model's estimate (\ref{coulomb_model})
exhibits instead a nuclear Coulomb energy of tens or a few hundred MeV
per nucleus (of which there are a few per molecule). Thus the true
$|\bm{\nabla}\bar\psi|$ is a fraction $< 10^{-7}({R/r})^3$ of the
Coulomb model prediction.  Hence, if it were possible to accurately
measure $\mathcal{T}$'s acceleration at $r\sim 10R$ from $\mathcal{S}$,
the ferroelectric would cause WEP violations at the fractional level
$<10^{-16}(l/\ell_P)^2$ or smaller.  This is within the projected
sensitivity of the STEP experiments (which, however, will not carry
aloft sources of $\bar\psi$).  But because the
\textit{electric} field near a ferroelectric is very strong, on the
order $10^8\,(R/r)^3\, \text{V}/\text{cm}$ for PbTiO$_3$, it would
probably prove necessary to work at considerably smaller
$R/r$ to avoid electric perturbations of $\mathcal{T} $.  The WEP
violation would then most surely become unobservable at the projected
STEP sensitivity.

To sum up this subsection, there are no clear cases where extant or
planned tests of the WEP are expected to detect violations connected
with $\alpha$ variability which originate in the electric structure of
matter.   But a STEP-like experiment based on a massive ferroelectric
source of $\bar\psi$ field in Earth's orbit would be a step closer for
detection of WEP violation arising from $\alpha$ variability.

\subsection{\label{sec:why}Why does the Coulomb model fail so badly ?}

Why is the Coulomb model recapitulated in Sec.~\ref{sec:naive}, with
its seemingly unassailable logic, so inaccurate ?  The answer to this
puzzle hinges on a peculiar cancellation which parallels that which
facilitated the passage from the complicated Eq.~(\ref{force0}) to the
simple Eq.~(\ref{force'}) in the context of one charged particle.  I
now explain with the help of macroscopic averaging.

In $\mathcal{S}$  each microscopic particle's mass is subject to
Eq.~(\ref{dmi}).  Therefore, macroscopic averaging of
Eq.~(\ref{psieqn'}) for $\psi$ entails the replacement
\begin{widetext}
\begin{equation}
\sum_{i} {\partial m_i
c^2\over\partial\psi}\delta^3(\mathbf{x}-\mathbf{z}_i)\rightarrow
-{1\over \mathcal{V}}\int_{\mathcal{V}} d^3x
\sum_{i\in
\mathcal{V}}\kappa^{-1}e_{0i}\tan(\kappa\Phi(\mathbf{z}_i))
\delta^3(\mathbf{x} -\mathbf{z}_i)\approx -{1\over\mathcal{V}}\sum_{i\in
\mathcal{V}} e_{0i}\Phi(\mathbf{z}_i)
\label{sigma1}
\end{equation}
\end{widetext}  where the approximation uses the fact
(Sec.~\ref{sec:size}) that at the microscopic level
$\kappa|\Phi|\ll 1$, even near a charge, to discard terms of
$\mathcal{O}(\kappa^3\Phi^3)$.   Note that this expression is
\textit{minus} twice the formal microscopic Coulomb energy density in
$\mathcal{S}$; the sign is opposite that expected from the logic of the
Coulomb model.

Next one needs the macroscopic average of
$e^{-2\psi}\mathbf{E}^2$ in
$\mathcal{S}$.  In the above mentioned approximation
Eq.~(\ref{newelec}), $\mathbf{E}=-e^{2\psi} \bm{\nabla}\Phi$, gives with
Eq.~(\ref{newpsi}), $e^{2\psi}\approx 1+\kappa^2
\Phi^2+\cdots\ $, that
\begin{widetext}
\begin{equation} {1\over 4\pi}e^{-2\psi}\mathbf{E}^2\rightarrow {1\over
4\pi
\mathcal{V}}\int_\mathcal{V} d^3 x
[(\bm{\nabla}\Phi)^2+\kappa^2\Phi^2\mathbf{E}^2] =
{1\over\mathcal{V}}\sum_{i\in \mathcal{V}} e_{0i}\Phi(\mathbf{z}_i)
-{1\over 4\pi
\mathcal{V}}\left[\oint_{\partial
\mathcal{V}}\Phi\mathbf{E}\cdot d\mathbf{s}-\kappa^2\int_\mathcal{V} d^3
x\,\Phi^2\mathbf{E}^2\right]
\label{sigma2}
 \end{equation}
\end{widetext} The second form comes from integrating the
$(\bm{\nabla}\Phi)^2$ with the help of Gauss's theorem and employing
Eq.~(\ref{Poisson}).

Note that the aforementioned microscopic Coulomb energy density term
cancels out from the combined contributions (\ref{sigma1}) and
(\ref{sigma2}).  One term that survives is the \textit{surface}
integral over $\Phi\mathbf{E}$.  To estimate it recall that if
$\partial\mathcal{V}$ is pushed outward somewhat, $\Phi$ on it becomes
largely immune to fluctuations from individual charges composing
$\mathcal{S}$.  Roughly, then, by Gauss's electric law the surface term
in Eq.~(\ref{sigma2}) is,
$\mathcal{V}^{-1}\langle\Phi\rangle\sum_{i\in
\mathcal{V}} e_{0i}$ where $\langle\Phi\rangle$ is an average of $\Phi$
over the surface. But this expression is on the order of the Coulomb
energy per unit volume associated with the
\textit{net free} charge contained in $\mathcal{V}$.  And we have seen
by example in Sec.~\ref{sec:lab} that this last energy is much smaller
than its microscopic counterpart
$\mathcal{V}^{-1}\sum_{i\in
\mathcal{V}} e_{0i}\Phi(\mathbf{z}_i)$.

What about the $\kappa^2\Phi^2\mathbf{E}^2$ integral in
Eq.~(\ref{sigma2})?  Let $L$ be the smallest microscopic length scale
on which $\Phi$ varies, e.g. $10^{-13}$ cm if we think of nucleons as
the smallest constituents, and perhaps
$10^{-17}$ cm if quarks and electrons are considered in their stead
(already in orthodox quantum electrodynamics, their electric fields are
not scale-free Coulombic).  Then one expects the maximum
$|\Phi|$ to be no larger than a unit charge divided by $L$.  Hence on
the the average  $\kappa^2\Phi^2\mathbf{E}^2/4\pi$ is bounded by
$\alpha(l/L)^2$ times the average microscopic Coulomb energy density
$\mathbf{E}^2/4\pi$.  This source of
$\bar\psi$ is a fraction $< 10^{-34}(l/\ell_P)^2$ of that assumed in
the Coulomb model.  It is thus amply clear how the cancellation of
Coulomb energy source terms in the equation for $\bar\psi$ causes the
Coulomb model to grievously overestimate
$|\bm{\nabla}\bar\psi|$.

\section{\label{sec:magnetic}Magnetic structure and WEP}

In ordinary matter, the next largest energy to Coulomb energy is energy
of the magnetic dipoles associated with spin and orbital angular
momentum of nuclei and electrons. Magnetic energy has so far been
ignored here. Eq.~(\ref{psieqn'}) is missing the magnetic part of the
term
$e^{-2\psi}f_{\mu\nu}f^{\mu\nu}$ in the original scalar equation
(\ref{wave_equation}).  And in Sec.~\ref{sec:why} magnetic
contributions have been left out of the mass term in
Eq.~(\ref{sigma1}).  Now that it is clear that Coulomb contributions
are well nigh irrelevant as sources of
$\bar\psi$, it is mandatory to take into account the purely magnetic
contributions.

\subsection{\label{sec:monopole}Magnetic monopoles: a shortcut to
dipoles}

It is a hard task to directly find the analog of the multicharge exact
solution (\ref{newpsi})--(\ref{newelec}) for a collection of magnetic
dipoles.  But as the example of Sec.~\ref{sec:why} makes clear, this
last solution is essential to derive the correct macroscopic form of
the mass term in the source of Eq.~(\ref{psieqn'}) for $\psi$.  To
overcome the problem I first find the analog of
Eqs.~(\ref{newpsi})-(\ref{newelec})  for a collection of magnetic
\textit{monopoles} of strengths $g_{0i}$,
$i=1,2, \cdots\ 2N$, and then let pairs of equal and oppositely charged
monopoles merge to form magnetic dipoles.  The monopoles serve as a
calculational crutch here and in the next subsection; they disappear
from the final results.

With the notation $\mathbf{B}\equiv \{f^{23},f^{31},f^{12}\}$ I first
replace Eqs.~(\ref{Phieqn'})--(\ref{psieqn'}) by
\begin{eqnarray}
\bm{\nabla}\times\left(e^{-2\psi}\mathbf{B}\right)=0 \hskip 4.8cm
\label{ampere}
\\
\bm{\nabla}\cdot\mathbf{B} =4\pi
\sum_i g_{0i}\delta^3(\mathbf{x}-\mathbf{z}_i)  \hskip 3.3cm
\label{gauss}
\\
\nabla^2\psi= 4\pi\kappa^2\left[\sum_i {\partial m_i
c^2\over\partial\psi}\delta^3(\mathbf{x}-\mathbf{z}_i)-{1\over
4\pi}e^{-2\psi}\mathbf{B}^2\right]
\label{psinew}
\end{eqnarray}  Eq.~(\ref{ampere}) is the space-space component of the
Maxwell-type Eq.~(\ref{Maxwell}) in a local Lorentz frame and with
current and time derivatives set to zero.  Eq.~(\ref{gauss}) does not
come from the action (\ref{SM}), but rather generalizes the Gauss law
$\bm{\nabla}\cdot\mathbf{B}=0$ (which follows -- also in variable
$\alpha$ theory --- from the representation of
$\mathbf{B}$ as a curl) to the case that there are magnetic monopoles
present. This is the argument for
\textit{not} including a factor $e^{-2\psi}$ in (\ref{gauss}).
Finally, Eq.~(\ref{psinew}) is the scalar equation
(\ref{wave_equation}) in a local Lorentz frame with electric field and
time derivatives dropped.

It is not possible here, in light of Eq.~(\ref{gauss}), to use a vector
potential for $\mathbf{B}$; however, it can be written in terms
of a scalar potential (its force on a monopole should be
conservative):
\begin{equation}
\mathbf{B}=-\bm{\nabla}\Psi.
\label{mag_field}
\end{equation}    It follows from Eq.~(\ref{gauss}) and the reasonable
boundary condition
$\Psi(\mathbf{x})\rightarrow 0$ as $|\mathbf{x}|\rightarrow
\infty$ that
\begin{equation}
\Psi(\mathbf{x})=\sum_i {g_{0i}\over |\mathbf{x}-\mathbf{z}_i|}
\label{Phi_tilde}.
\end{equation}  Eq.~(\ref{ampere}) now gives
$e^{-2\psi}\bm{\nabla}\psi\times\bm{\nabla}\Psi=0$ which uniquely
implies that $\psi=\psi(\Psi(\mathbf{x}))$.  Finally, taking the
Laplacian of this last relation gives
\begin{eqnarray}
\nabla^2\psi&=&\psi'\nabla^2 \Psi+\psi''(\bm{\nabla}\Psi)^2
\nonumber
\\ &=& -4\pi\psi'\sum_i g_{0i}\delta^3(\mathbf{x}-\mathbf{z}_i)  +\psi''
\mathbf{B}^2
\label{nabla'}
\end{eqnarray}

Comparison of  the second line of Eqs.~(\ref{nabla'}) and
Eq.~(\ref{psinew}) gives
\begin{eqnarray}
\psi'' &=& -\kappa^2e^{-2\psi}
\\
\kappa^2(\partial m_ic^2/\partial\psi) &=&-g_{0i}\psi'\quad
\textrm{at }\quad
\mathbf{x}=\mathbf{z}_i.
\label{dm2}
\end{eqnarray} The first of these integrates to
\begin{equation}
\psi'^2=\kappa^2(e^{-2\psi}+\tilde\varpi)
\label{psi'2}
\end{equation} with $\tilde\varpi$ a constant.  All the solutions of
this equations may be obtained as in Appendix~\ref{sec:solutions}:
\begin{equation} e^\psi =\left\{ \matrix{
\pm(\kappa \Psi+\tilde\chi); & \tilde\varpi=0, \cr
\\
\pm \tilde\varpi^{-1/2}\text{sinh}( \tilde\varpi^{1/2}\kappa
\Psi+\tilde\chi); &
\tilde\varpi>0, \cr
\\ |\tilde\varpi|^{-1/2}\cos(|\tilde\varpi|^{1/2}\kappa
\Psi+\tilde\chi); & \tilde\varpi<0
\cr } \right.
\label{summary_mag}
\end{equation}  with $\tilde\chi$ a second integration constant.

Because we made no guesses or \textit{ansatze} along the way to
Eqs.~(\ref{summary_mag}), these last together with
Eq.~(\ref{mag_field}) and (\ref{Phi_tilde}) must exhaust the solutions
of Eqs.~(\ref{ampere})--(\ref{psinew}) with boundary conditions
$\Psi(\mathbf{x})\rightarrow 0$ as $|\mathbf{x}|\rightarrow 0$.  In
this connection it is appropriate to remark that the substitution
$e^{2\psi}\mathbf{B}\mapsto\mathbf{E}$ followed by
$\psi\mapsto -\psi$ and $g_{0i}\mapsto e_{0i}$, transform the magnetic
solution~(\ref{mag_field})--(\ref{Phi_tilde}) together with each of the
five variants in Eq.~(\ref{summary_mag}) into one of the five variants
of the electric solution (\ref{field})-(\ref{choices}) of
Sec.~\ref{sec:derive}.  A similar transformation maps each electric
solution back onto a magnetic solution. The one-to-one correspondence
between magnetic and electric solutions reflects duality symmetry of
the $\alpha$ variability framework.  Since all time-independent
magnetic solutions have been found, the principle of duality assures us
that all time-independent electric solutions have been found as well.

Also by duality, the magnetic solution with $\tilde\varpi<0$, which is
dual to the physical electric solution ($\varpi<0$), must be regarded
as the physical choice. In fact, if invoking by now well worn
arguments, we set
$\tilde\varpi=-1$ and $\tilde\chi=0$, that solution is consistent with
the reasonable boundary condition
$\psi\rightarrow 0$ as
$|\mathbf{x}|\rightarrow \infty$, whereas the solutions with
$\tilde\varpi\geq 0$ cannot be made so.  Thus the unique physical
solution for a collection of magnetic monopoles at rest is given by
Eqs.~(\ref{mag_field}) together with
\begin{eqnarray} e^\psi&=&\cos(\kappa\Psi)
\label{psi2}
\\
\mathbf{B}&=&-\bm{\nabla}\Psi.
\label{newmag}
\end{eqnarray}

\subsection{\label{sec:dipole}Magnetic dipoles as source of
$\bar\psi$}

Returning to our main subject, magnetic dipole energy, I imagine that
the monopole labeled by $i+N$ has magnetic charge
$-g_{0i}$, and denote the vector from monopole $i+N$ to monopole $i$ by
$\mathbf{d}$.  Then if as the pair are allowed to approach each other
adiabatically, $|g_{0i}|$ is made to grow in such a way that
$g_{0i}\mathbf{d}\rightarrow
\bm{\mu}\neq 0$ in the limit, a point magnetic dipole is formed.  It
possesses magnetic moment
$\bm{\mu}$ and may be labeled by $i$.  I still refer to its position as
$\mathbf{z}_i$.

The total magnetic potential is
\begin{eqnarray}
\Psi&=&\lim_{|\mathbf{d}|\rightarrow 0}\sum_{i=1}^N
\left[{g_{0i}\over|\mathbf{x}-\mathbf{z}_i-\mathbf{d}/2|}+
{-g_{0i}\over|\mathbf{x}-\mathbf{z}_i+\mathbf{d}/2|}\right]
\nonumber
\\ &=& \sum_{i=1}^N{\bm{\mu}_i\cdot (\mathbf{x}-\mathbf{z}_i)\over
|\mathbf{x}-\mathbf{z}_i|^3}.
\label{mag_dip}
\end{eqnarray} Thus $e^{\psi}$ is explicitly known from
Eq.~(\ref{newmag}).  Performing the gradient in Eq.~(\ref{newmag})
gives for the field of the dipole,
\begin{equation}
\mathbf{B}=\sum_{i=1}^N\left[{3\bm{\mu}\cdot(\mathbf{x}-\mathbf{z}_i)\
(\mathbf{x} -\mathbf{z}_i)\over
|\mathbf{x}-\mathbf{z}_i|^5}-{\bm{\mu}\over
|\mathbf{x}-\mathbf{z}_i|^3}\right].
\label{dipole_field}
\end{equation} This coincides (apart from the singularities at
$\mathbf{x}=\mathbf{z}_i$) with the Maxwellian expression for the field
of an array of magnetic dipoles derived from the vector
potential~\cite{jackson}.  It seems reasonable that were magnetic
dipoles to be represented by tiny current loops, the field would still
be (\ref{dipole_field}). Eqs.~(\ref{psi2})--(\ref{mag_dip}) thus
represent the multimagnetic dipoles solution within the framework.

What is the source term in the equation for $\psi$ for matter made
exclusively of magnetic dipoles ?  From Eq.~(\ref{psi2}) it follows that
\begin{equation}
\nabla^2 \psi = \kappa
\tan(\kappa\Psi)\bm{\nabla}\cdot\mathbf{B}-
\kappa^2\sec^2(\kappa\Psi)\mathbf{B}^2
\end{equation} It is easy to verify, for example by applying Gauss'
theorem to Eq.~(\ref{dipole_field}), that
$\bm{\nabla}\cdot\mathbf{B}$ vanishes everywhere, and has no
$\delta$-function singularity at the position of a dipole. Since
$\sec^2(\kappa\Psi)=e^{-2\psi}$, comparison of the last result with
Eq.~(\ref{psinew}) shows that the mass-dependent source term is absent
for magnetic dipoles.  It is easy to get a wrong answer in this
respect, for example, by trying to combine the mass-dependent terms of
the positive and negative poles \textit{before} taking the limit
$|\mathbf{d}|\rightarrow 0$. Since the mass-dependent source term would
arise naturally from the action, the conclusion must be that the mass
of a pointlike magnetic dipole, unlike that of a pointlike charge, does
not depend on $\psi$.

An immediate consequence is that a magnetic dipole moment in an
exterior electromagnetic field is not subject to an anomalous force of
the sort appearing for an electric charge, cf. Eq.~(\ref{force0}).
The only force would be the usual
$(\bm{\mu}\cdot\bm{\nabla})\mathbf{B}$ obtainable in our context by
combining the forces on two magnetic poles.  All this means  that no
violations of the WEP are expected in the motion of matter with pure
magnetic dipole structure; it is unnecessary in this connection to
carry out the analog of the calculation in Secs.~\ref{sec:momentum} and
\ref{sec:WEP}.

But does a source $\mathcal{S}$ with purely magnetic dipole structure
cause WEP violations in the motion of a test body
$\mathcal{T}$ made up of electric charges ?   In analogy with the
argument in Sec.~\ref{sec:psi} it follows from Eq.~(\ref{psi2}) that
\begin{equation}
\bar\psi\approx -{\scriptstyle 1\over\scriptstyle
2}\kappa^2\overline{\Psi^2}=-{\scriptstyle 1\over\scriptstyle
2}\kappa^2\bar {\Psi}^2  - {\scriptstyle 1\over\scriptstyle
2}\kappa^2\overline{(\Psi-\bar\Psi)^2}
\label{barpsi2}
\end{equation} Contrary to what happens in the presence of a source
with electric structure, cf. Sec.~\ref{sec:psi}, here the physical
$\alpha$ has its value slightly
\textit{depressed} in relation to the asymptotic value.  As in
Sec.~\ref{sec:psi},
\textit{outside} a macroscopic source $\mathcal{S}$, the fluctuation
term should be relatively small compared to $\bar
\Psi^2$, except in rather artificial situations where
$\bar\Psi$ vanishes ($\mathcal{S}$ has no net magnetic moment).  The
latter are not important in our context, so henceforth I shall drop the
fluctuation and write just
$\Psi$ for $\bar\Psi$.

Supposing $\mathcal{S}$ to be a sphere with uniform magnetization
$\mathbf{M}$, $\Psi$ in Eq.~(\ref{mag_dip}) is well approximated at
distance
$r\gg R$ from the sphere's center by $(4\pi R^3/3)\mathbf{M}\cdot
\mathbf{r} r^{-3}$.  Then
\begin{equation} |\bm{\nabla}\bar\psi|\approx 4\kappa^2 ({R/r})^3[(4\pi
R^3/3)
\mathbf{M}^2]r^{-2}
\label{gradpsi}
\end{equation}  If Earth is the source $\mathcal{S}$, its crudely
dipolar magnetic field ($\sim 0.25$ G at the magnetic poles) allows
its representation by a uniformly magnetized sphere with
$|\mathbf{M}|\sim 3\times 10^{-2}$ cgs~\cite{jackson}.  The energy
represented by the square brackets in Eq.~(\ref{gradpsi}),
$\sim 9.7\times 10^{23}\, \text{erg}$, is a fraction $\sim 10^{-22}$ of
the corresponding factor in Eq.~(\ref{coulomb_model}) for the Coulomb
model (I take
$\zeta=10^{-3}$ for Earth).  Hence Earth, by virtue of its magnetic
structure, causes WEP violations at the fractional level $10^{-28}
(l/\ell)^2$, well beyond anything measurable in the foreseeable future.

Another example of $\mathcal{S}$ is furnished by a ferromagnetic
sphere.  Iron, one of the ferromagnets with the highest saturation
magnetization, can reach
$|\mathbf{M}|\approx 1.7\times 10^3
\,\text{cgs}$~\cite{kittel}.  The energy in the square brackets in
Eq.~(\ref{gradpsi}) is then $\approx 1.2\times 10^7
\,\text{erg}/\text{cm}^3\, R^3$.  Being iron ($\rho=7.87\,
\text{g}/\text{cm}^3$ and mass number 55.84) the sphere contains $\sim
3.55\times 10^{23}/\text{cm}^3\, R^3$ Fe nuclei; each contributes $\sim
125$ MeV of Coulomb energy~\cite{segre} for a total of $\sim 7.11\times
10^{19}\
\text{erg}/\text{cm}^3\ R^3$.   Thus the true
$|\bm{\nabla}\bar\psi|$, Eq.~(\ref{gradpsi}), amounts to a fraction
$1.7\times 10^{-13} (R/r)^3$ of the  Coulomb model prediction. It
follows that a ferromagnetic source of
$\bar\psi$ in orbit would cause WEP violations at a fractional level
well below $10^{-19}$.  This is beyond the projected STEP sensitivity
(and, of course, the STEP experiment will not carry the equivalent of
$\mathcal{S}$).

To be sure the above procedure is a bit cavalier. The exact solution
(\ref{psi2})--(\ref{mag_dip}) which forms the ultimate basis of the
calculation contains no electric charges, while Earth and iron have
plenty of them.   But this defect is not serious.  Let $\psi\equiv
\psi_{\text{e}}(\Phi)+\psi_{\text{m}}(\Psi)$ with
$\psi_{\text{m}}$ the function $\psi(\Psi)$ in Eq.~(\ref{psi2}) and
$\psi_{\text{e}}$ the $\psi(\Phi)$ defined by Eq.~(\ref{newpsi}) for
the collection of electric charges.  Then the electric field
$\mathbf{E}$ in Eq.~(\ref{newelec}) solves the Gauss equation
(\ref{Phieqn'}), to within a relative correction of
$\mathcal{O}(\psi_{\text{m}})$, and is a  gradient.  Likewise, the
magnetic field $\mathbf{B}$ [Eq.~(\ref{dipole_field})] deriving from
$\Psi$ in Eq.~(\ref{mag_dip}) satisfies, to within a fractional
correction of $\mathcal{O}(\psi_{\text{e}})$, the Ampere-like
Eq.~(\ref{ampere}), and is also divergence free (Gauss's magnetic
equation).

Further, adding Eq.~(\ref{psieqn'}) for $\psi_{\text{e}}$ to
Eq.~(\ref{psinew}) (without the mass term) for
$\psi_{\text{m}}$ gives, to within fractional corrections of
$\mathcal{O}(\psi_{\text{e}})$ and
$\mathcal{O}(\psi_{\text{m}})$, the correct equation for the full
$\psi$ relevant for a collection of charges \textit{and} magnetic
dipoles, cf. Eq.~(\ref{wave_equation}):
\begin{equation}
\nabla^2\psi= 4\pi\kappa^2\Big[\sum_i {\partial m_i
c^2\over\partial\psi}\delta^3(\mathbf{x}-\mathbf{z}_i)+{1\over
4\pi}e^{-2\psi}(\mathbf{E}^2-\mathbf{B}^2)\Big]
\label{last}
\end{equation}  The sum here extends only over electric charges.
Plainly the electric and magnetic contributions to
$\psi$ are additive (because $|\psi_{\text{e}}|\ll 1$ and
$|\psi_{\text{m}}|\ll 1$).  Thus the correct $\bar\psi$ to use for real
matter (charges plus magnetic dipoles) is, not that in
Eq.~(\ref{barpsi2}), but rather
\begin{equation}
\bar\psi\approx {\scriptstyle 1\over\scriptstyle
2}\kappa^2(\bar{\Phi}^2 -\bar{\Psi}^2).
\label{reallylast}
\end{equation}

In view of this composition rule, the results of Sec.~\ref{sec:dipole}
that WEP violations originating from
$\psi_{\text{m}}$ will be unobservable for the foreseeable future,
together with those of Sec.~\ref{sec:psi} that effects coming from
$\psi_{\text{e}}$ have the same status, show there are no clear cases
where extant or planned tests of the WEP are expected to detect
violations connected with $\alpha$ variability.  Still outstanding is
the question whether motion of charges or dipoles in matter  --- totally
ignored in this paper --- could provide a loophole from this conclusion.

\section{\label{sec:real}Resolving  a cosmological conundrum}

The cosmological evolution of $\alpha$ according to the framework is
influenced by the nature of $\psi$'s source and by a certain
integration constant $t_C$ (a cosmological time scale)~\cite{bek}.  If
the source is  described with the Coulomb model applied to ordinary
matter of cosmological abundances, and the appealing choice
$t_C=0$ is made, then $\alpha$ is predicted to
\textit{decrease} throughout the matter dominated
era~\cite{bek,olive,SBM}.  The claimed cosmological rise of
$\alpha$~\cite{webb} thus led Olive and Pospelov~\cite{olive} and
Barrow,  Sandvik and Magueijo~\cite{BSM2} to conclude that the scalar
field is coupled  mostly to dark matter with a coupling opposite in
sign to the naive one.  To quote Ref.~[\onlinecite{BSM2}], `` ... the
dark matter constituents have to have a high magnetostatic energy
content (one possible contender would be superconducting cosmic strings
which have $\zeta\approx -1$)''; as clear from Eq.~(\ref{last}),
magnetic fields are sources of $\psi$ with effectively negative $\zeta$.

Arguably the mentioned resolution of the cosmological conundrum amounts
to trading one cosmic mystery (growing
$\alpha$) for another (dark matter with especially unusual
electromagnetic properties).  A much more natural solution is offered
by the realization (see Sec.~\ref{sec:why}) that the Coulomb energy
cancels from the source term of $\psi$'s equation.  This should also be
true for cosmological baryonic matter because one can think of each
small region of the cosmological medium as being at rest in a local
Lorentz frame.  Consequently, the source of $\psi$'s cosmological
evolution is principally the $\mathbf{B}^2$ term contributed by
matter's magnetic dipoles to the cosmological version of
Eq.~(\ref{last}). [By Sec.~\ref{sec:dipole} the mass term in
Eq.~(\ref{wave_equation}) is absent for magnetic dipoles, thermal
radiation does not contribute at all because
$\mathbf{E}^2-\mathbf{B}^2$ vanishes for electromagnetic
radiation~\cite{bek}, and cosmological magnetic fields are too weak to
make a difference.]  Thus the switch in sign of the source term
required by the claimed cosmological increase of
$\alpha$ comes about automatically by considering the magnetostatic
energy of baryonic cosmological matter.   As I show now, the rate at
which $\alpha$ grows also comes out of the right order without special
assumptions.

For cosmological $\alpha$ variability the interesting field is
$\bar\psi$.  As already discussed, it will obey an equation like
(\ref{wave_equation}) but with the source averaged over a macroscopic
region. This average magnetic energy density is easiest evaluated by
first computing the magnetic energy associated with a given mass of
cosmological matter.

Now in baryonic matter the magnetic energy is principally tied to the
protons' and electrons' magnetic moments ($^4$He nuclei have no
magnetic moment).  A lower bound on (and a passably good estimate for)
the magnetic energy $E_p$ of the proton should result from regarding it
as a perfect magnetic dipole and ignoring the energy interior to the
proton radius
$R_p$.  Integrating $\mathbf{B}^2/8\pi$ (cf. the electric energy in
Appendix~\ref{sec:corrections}) with
$\mathbf{B}$ in Eq.~(\ref{dipole_field}) over the space outside
$R_p$ gives
$E_p\agt \mu^2/3R_p{}^3$.  Now for the proton $\mu = 2.98
\times e\hbar/2m_p c$~\cite{segre}, whereas $R_p\approx 5.7 (\hbar/m_p
c)$.  Thus $E_p\agt 2.54\times 10^{-5} m_p c^2$.  For the electron
$\mu=e\hbar/2m_e c$.  In view of the
\textit{Zitterbewegung} phenomenon is seems reasonable to integrate the
magnetic energy of the dipole field only down to the Compton length
$\hbar/m_e c$.  Doing this as in
 the proton calculation gives $E_e\agt 6.1\times 10^{-4}m_e c^2$.
Other magnetic energies, e.g. those connected with spin-orbit and
hyperfine splittings, are small on the Bohr scale $\alpha^2 m_e c^2$,
and thus negligible here.

In the context of a matter-dominated Robertson-Walker expanding model
(with no cosmological constant and neglecting
$\psi$'s effect on the expansion as appropriate at late cosmological
time), one can use Eq.~(31) of Ref.~[\onlinecite{bek}] for the rate of
change of $\alpha$:
\begin{equation}
\dot \alpha/\alpha = -(3/4\pi)(l/\ell_P)^2\zeta\,\Omega_b H{}^2 t
\label{dotalpha}
\end{equation} Here $H$ is the Hubble ``constant'' at cosmological time
$t$,
$\Omega_b$ the baryon density parameter and $|\zeta|$ the fraction of
the cosmic baryon mass density $\rho_b$ which acts as source of
$\bar\psi$ (here $\zeta<0$). Now 23\% of $\rho_b$ consists of
$^4$He. From the above results the magnetic energy
\textit{density} from helium's two electrons, and the protons and
electrons from hydrogen, when appropriately weighted by their cosmic
abundances, is $\mathcal{E}_m\agt 1.98\times 10^{-5}\rho_b c^2$,
meaning that $\zeta\alt - 1.98\times 10^{-5}$.  With
$Ht=2/3$ (matter domination) and today's favorite value
$\Omega_b=0.03$, Eq.~(\ref{dotalpha}) predicts
\begin{equation}
\dot \alpha/\alpha \agt 6\times 10^{-8} (l/\ell_P)^2\, t^{-1}.
\end{equation}

As advertised, $\alpha$ is predicted to increase with time.  In order
for its overall fractional change over the last
$10^{10}$ y to match the observed $0.7\times 10^{-5}$, it is necessary
that $l\alt 10 \ell_P$.  This is  a reasonable value consistent with
the framework's assumptions and with all tests of the WEP (see
Secs.~\ref{sec:psi} and
\ref{sec:dipole}).  Hence nothing but baryonic matter is necessary to
explain the claimed cosmological sense and rate of $\alpha$
variation.   Were this last to be ruled out by future observations, one
would have to conclude that $l$ is smaller than mentioned.  Within the
spirit of the framework  an $l$ much below $\ell_P$ is unacceptable,
and so one can risk a flat prediction that in recent epochs $t\dot
\alpha/\alpha > 10^{-8}$.  Observational exclusion of this bound would
be tantamount to certifying that standard Maxwellian electrodynamics
(the case $l\equiv 0$) is the exact classical description of
electromagnetism.

\section{\label{caveats}Conclusions and Caveats}

The issue of $\alpha$ variability has long been connected with the
possibility of weak equivalence principle violations coming from the
classical Coulomb energy contribution to particle masses.  The central
conclusion of this paper is that in the general framework for $\alpha$
variability described in Ref.~[\onlinecite{bek}], a compensating mechanism
exists which prevents such violations from being measurable, at least
in the foreseeable future.   The treatment has been classical or
tree level.  Quantum considerations raise a potential problem for this
kind of theory.

Banks, Dine and Douglas~\cite{bdd} have argued that in a theory where a
scalar field couples to the electromagnetic scalar
$f_{\mu\nu}f^{\mu\nu}$, as in the action (\ref{SM}), the residual
vacuum energy of matter fields is very much larger than the observed
cosmological constant unless the parameters are finely tuned.  Recall
that even with no scalar coupling the vacuum energy is known to be
formally very large.  But it is widely believed that some as yet ill
understood mechanism nearly cancels it.  Ref.~[\onlinecite{bdd}] points
out that because of the variation of the scalar field required for the
suggested explanation of cosmological $\alpha$ variation to work, this
cancellation can be successful only over a very short interval of
cosmological time, with the residual vacuum energy becoming intolerably
large earlier and thereafter.

In light of this argument, can a theory like the framework used here be
taken seriously ?  One has to recall that the aforesaid criticism
presupposes the existence of a cancellation of the vacuum energy by a
mechanism whose nature is not generally agreed upon today.  One can
well imagine that when that mystery is lifted, a resolution for the
problem of large variation of the residual vacuum energy with the
scalar field might also become apparent.  At any rate, it is premature
to draw any final conclusions at this stage since the problem remarked
upon in Ref.~\onlinecite{bdd} is an integral part of the unsolved
mystery of the small cosmological constant.

\begin{acknowledgments} I am indebted to Cristophe Grojean, John
Moffat, Jonathan Oppenheim and especially to Gia Dvali and Matias
Zaldarriaga for remarks and criticisms which sharpened my understanding
of the issues discussed here.  This research was supported by grant No.
129/00-1 of the Israel Science Foundation.
\end{acknowledgments}

\appendix

\section{\label{sec:solutions}Solutions for $\psi$}

For $\varpi=0$ take a square root of Eq.~(\ref{psi'}) to get
\begin{equation} e^{-\psi}\psi'=\pm \kappa
\end{equation} which may immediately be integrated to
\begin{equation} e^\psi=\mp(\kappa V+\chi)^{-1}
\end{equation} with $\chi$ a constant.

For $\varpi\neq 0$ set $\psi= \ln y+{\scriptstyle 1\over\scriptstyle
2}\ln|\varpi|$.  Then after taking a square root, Eq.~(\ref{psi'})
transforms into
\begin{equation} {dy\over y\sqrt{y^2+1\cdot\text{sgn}(\varpi)}}=\pm
|\varpi|^{1/2}\kappa\, d V
\label{diff}
\end{equation} In the case $\varpi>0$ the l.h.s. of this last equation
is the differential of $\text{arccsch}(y)$ [the upper (lower) signs
corresponding to negative (positive)
$y$].  Thus
\begin{equation} y=\pm \text{csch}(\varpi^{1/2}\kappa V+\chi).
\end{equation}

In the case $\varpi<0$ the l.h.s. of Eq.~(\ref{diff}) is the
differential of $\text{arcsec}(y)$ for either sign of $y$.   Thus since
$\sec(\pm y)=\sec (y)$,
\begin{equation} e^\psi=\left\{ \matrix{
\pm(\kappa \Phi+\chi)^{-1}; & \varpi=0, \cr
\\
\pm \surd\varpi\,\text{csch}( \surd\varpi\,\kappa
\Phi+\chi); &
\varpi>0, \cr
\\
 \surd|\varpi|\,\sec(\surd|\varpi|\,\kappa \Phi+\chi); &
\varpi<0. \cr }
\right.
\label{summary}
\end{equation}

\section{Applicability of the Reissner-Nordstr\" om
metric}\label{sec:corrections}

Here I focus on an elementary charge.  To assess the corrections to the
metric coming from the scalar field I write down the energy-momentum
tensor contributions from
$\psi$ and $f^{\mu\nu}$ in curved spacetime in accordance with
Eqs.~(\ref{Spsi})--(\ref{SM}):
\begin{eqnarray} T_{(\psi)}{}^\mu{}_\nu&=&{1\over
4\pi\kappa^2}\Big(\psi_,{}^\mu
\psi_{,\nu} -{\scriptstyle 1\over \scriptstyle 2} g^\mu{}_\nu
\psi_,{}^\alpha \psi_{,\alpha}\Big)
\label{T_psi}\\ T_{(f)}{}^\mu{}_\nu&=&{e^{-2\psi}\over
4\pi}\Big(f^{\mu\alpha}f_{\nu\alpha}-{\scriptstyle 1\over
\scriptstyle 4} g^\mu{}_\nu f^{\alpha\beta}f_{\alpha\beta}\Big)
\label{T_Maxwell}
\end{eqnarray} In either tensor energy densities and stresses have the
same magnitudes for a spherical static solution for which necessarily
$\psi_{,r}$ and $f^{tr}$ are the only surviving components of
$\psi_{,\mu}$ and $f^{\mu\nu}$, respectively). In the weak gravity
region $(r\gg R)$ one may use Eq.~(\ref{coulomb}), the fact that
$\bm{\nabla}\psi=\psi'\bm{\nabla}\Phi$ and Eq.~(\ref{soln''}) (with
$\chi=0$) to get for the said solution
\begin{equation}
{|T_{(\psi)}{}^\mu{}_\nu|\over|T_{(f)}{}^\mu{}_\nu|}=\kappa^{-2}\psi'{}^2
e^{2\psi}=\sec^2 \kappa\Phi\, \tan^2 \kappa\Phi
\end{equation}

$T_{(\psi)}{}^\mu{}_\nu$ is thus negligible compared to
$T_{(f)}{}^\mu{}_\nu$ all the way in to a radius where
$|\Phi|$ is no longer  small compared to $\pi^{3/2}(\hbar
c)^{1/2}l^{-1}$.  According to Sec.~\ref{sec:size} the said radius
represents the smallest possible extension of the charge.   Hence,
unless the charge is as compact as permitted, one can consistently
neglect $T_{(\psi)}{}^\mu{}_\nu$ in the above equations over the entire
charge's exterior, and can do so over most of the exterior in the
compact charge case.

A look at Eq.~(\ref{elecfield}) shows that $f_{tr}$ here is
$e^{2\psi}$ times its Maxwellian counterpart
$c\bm{\nabla}\Phi$.  Hence, $T_{(f)}{}^\mu{}_\nu$ here differs from the
pure Maxwellian energy momentum tensor also by a factor
$e^{2\psi}=\sec^2\kappa\Phi$, which is very close to unity all the way
in to the radius mentioned above.  Thus the Maxwellian energy momentum
tensor replaces the full
$T^\mu{}_\nu$ throughout the whole exterior of a not maximally compact
charge, and throughout most of it for a compact charge.  As a
consequence the Reissner-Nordstr\" om metric (\ref{metric}) is rather
accurate outside the charge.

\section{\label{sec:m}The mass function}

What can be said about spatial variation of masses of electrically
charged particles through their dependence on
$\psi$ ?  For simplicity of notation I focus on that of charge 1 in the
cluster discussed in Sec.~\ref{sec:momentum}. I first rewrite
Eq.~(\ref{dmi}) with the help of Eq.~(\ref{newpsi}) as
\begin{equation} (\partial m_1
c^2/\partial\Phi)=-e_{01}\tan^2(\kappa\Phi).
\label{dm'}
\end{equation} Evidently $\Phi$ here means the value of the potential
(\ref{Phi'}) evaluated at the position of charge
$1$, and of course this potential is large there (not infinite, though,
since as mentioned in Sec.~\ref{sec:momentum}, charges cannot be exact
points in this framework).   The equation refers to the {\it change\/}
of $m_1$ resulting from the local {\it change\/} of $\Phi$ which might
be due, for example, to the other charges being moved around
(adiabatically).

Proceeding purely formally I use the fact that $d\tan x/dx=1+\tan^2 x$
to integrate Eq.~(\ref{dm'}):
\begin{equation} m_1
c^2=m_{01}c^2+e_{01}[\Phi-\kappa^{-1}\tan(\kappa\Phi)]
\label{m}
\end{equation}  Here $m_{01}$ is a constant of integration.  It is
clear from this that $m_1$ is invariant under conjugation of {\it
all\/} charges, as would be expected by C invariance of the
electromagnetic interaction; although the framework's assumptions
require only T invariance, the theory's action
(\ref{Spsi})--(\ref{gaction}) is  actually C invariant as well.
According to Eq.~(\ref{soln''}),
$\kappa\Phi=  \text{arcsec}\, e^\psi$; since
$\tan^2(\kappa\Phi)=\sec^2(\kappa\Phi)-1$ one has
\begin{equation} m_1
c^2=m_{01}c^2+\kappa^{-1}e_{01}\left(\text{arcsec}\,
e^{\psi}-\sqrt{e^{2\psi}-1}\right).
\label{newm}
\end{equation} The appropriateness of choosing the shown sign of the
square root is verified by differentiating $m_1(\psi)$ and comparing
the result with Eqs.~(\ref{dmi}) and (\ref{soln''}).  The opposite sign
is excluded on this ground.

As anticipated, the space-dependent contribution to $m_1$ is not
proportional to $e_{0i}{}^2e^{2\psi}$ as would be assumed in the
Coulomb model.

Unfortunately, Eqs.~(\ref{m}) and (\ref{newm}) are not immediately
useful; they contain a contribution from the self potential of particle
1 or from its self $\psi$ field.  If there were no other charges in the
universe, one would expect
$m_1$ to subsume these contributions.  Absorbing the self terms of
Eq.~(\ref{m}) into $m_{01}$ gives
\begin{equation} m_1
c^2=m_{01}c^2+e_{01}\big\{\Phi_{i>1}-\kappa^{-1}\big[\tan(\kappa\Phi)
-\tan(\kappa\Phi_{i=1})\big]\big\}
\end{equation} The identity $(\tan x-\tan y)\cos x\cos y=\sin(x-y)$
allows us to rewrite this as
\begin{equation} m_1 c^2=m_{01}c^2+e_{01}[\Phi_{i>1}-\kappa^{-1}A
 \sin(\kappa\Phi_{i>1})]
\label{m01}
\end{equation} where $A\equiv
\sec(\kappa\Phi_{i=1})\sec(\kappa\Phi)$ is evaluated at particle 1's
center.

Quite generically $\kappa|\Phi_{i>1}|\ll 1$.  This merely requires that
the overall size $\tilde R$ of the cluster of charges $i=2, 3,
\cdots\,$, and its overall charge \textit{sans} $e_{01}$,
$\tilde Q$, satisfy
\begin{equation}
\tilde R\gg \pi^{-3/2}(\tilde Q^2 /\hbar c)^{1/2} (l/\ell_P)\ell_P
\end{equation}   which is easily met by nucleons, nuclei, ions and
macroscopic charged objects.  One expects charge 1 to be relatively
more compact so that
$|\Phi_{i=1}|\gg |\Phi_{i>1}|$.  In this case Taylor expanding
$A$ to first order in $\kappa\Phi_{i>1}$ as well as
$\sin(\kappa\Phi_{i>1})$ to second order converts Eq.~(\ref{m01})  into
\begin{widetext}
\begin{equation} m_1 c^2=m_{01} c^2 - e_{01}\sec^2(\kappa\Phi_{i=1})
\times[\sin^2(\kappa\Phi_{i=1})\Phi_{i>1}+
\kappa
\tan(\kappa\Phi_{i=1})\Phi_{i>1}{}^2+\mathcal{O}(\Phi_{i>1}{}^3)]
\nonumber
\label{m1}
\end{equation}
\end{widetext} With the possible exclusion of the ``point'' leptons and
quarks, even elementary particles are much more extended than the
bound~(\ref{R}); thus
$\kappa|\Phi_{i=1}|\ll 1$.   Hence the variable part of $m_1$ is
generally very small compared to the Coulomb potential energy of that
charge in the cluster, $e_{01}\Phi_{i>1}$, and of opposite sign.

\end{document}